\documentclass[10pt,twoside,twocolumn]{article}
\usepackage[margin=0.75in]{geometry}
\usepackage{cite}
\usepackage{amsmath,amssymb,amsfonts}
\usepackage{algorithmic}
\usepackage{graphicx}
\usepackage{textcomp}
\usepackage{wrapfig,colortbl}
\usepackage{booktabs}
\usepackage{xcolor}
\usepackage{multirow}
\usepackage{rotating}
\usepackage{float}
\usepackage{dblfloatfix} 
\usepackage{authblk}
\usepackage{titlesec}
\usepackage[hidelinks]{hyperref}
\hypersetup{
    colorlinks=true,
    linkcolor=blue,
    citecolor=blue,
    urlcolor=blue
}

\definecolor{abstractbg}{rgb}{1,0.969,0.914}
\def\BibTeX{{\rm B\kern-.05em{\sc i\kern-.025em b}\kern-.08em
    T\kern-.1667em\lower.7ex\hbox{E}\kern-.125emX}}

\AtBeginDocument{%
  \setlength{\abovedisplayskip}{\abovedisplayshortskip}%
  \setlength{\belowdisplayskip}{\belowdisplayshortskip}%
}
\begin{document}
\title{Sparse Delta Integration method for the calculation of spatiotemporal pressure
fields of arbitrary ultrasound transducer geometries\thanks{This work was funded by the
European Research Council (ERC Synergy Grant N $^\circ$ 101118744 - Project
NeuroSonoGene).}\thanks{Code used for this article has been deposited at the Zenodo
repository (link).}}
\author{Deyver E. Rivera}
\author{Charlie Demené}
\author{Mickael Tanter}
\affil{Institute Physics for Medicine Paris, Inserm U1273, ESPCI Paris, PSL University,
CNRS UMR 8063, Paris, France \\ \texttt{mickael.tanter@espci.fr}}
\date{}

\twocolumn[
\begin{@twocolumnfalse}
\maketitle
\begin{abstract}
Accurate and efficient simulation of ultrasound pressure fields and pulse-echo responses
is essential for transducer design, beamforming optimization, and model-based imaging
research. Conventional Spatial Impulse Response (SIR) methods compute acoustic fields 
by explicitly sampling trapezoidal impulse
responses for each rectangular aperture subdivision, resulting in substantial
computational cost for large apertures, dense spatial grids, and high sampling
frequencies. In this work, we introduce Sparse Delta Integration (SDI), a new
mathematical framework for the far-field SIR of rectangular apertures that reformulates
the trapezoidal spatial impulse response as the double integration of a sparse set of
Dirac delta distributions.This reformulation provides compact expressions for both time-
and frequency-domain SIRs and enables vectorized implementations whose computational
cost is independent of the trapezoid duration, which previously varied for each
patch–field-point pair. We further derive Spectral SDI, an exact frequency-domain
formulation for pulse-echo simulations that removes part
of the conventional Fourier-domain convolution pipeline. Implemented in the open-source
Python package eSDIva, these methods achieve speedups of up to 180x incomputation of 
temporal SIR used for pressure-field simulations and up to 20x for pulse-echo RF
simulations relative to Field II, while maintaining high numerical accuracy with mean
squared errors below 10-5 and Pearson correlation coeffients close to unity.
\end{abstract}
\vspace{1.5em}
\end{@twocolumnfalse}
]

\section{Introduction}
\label{sec:introduction}

Accurate and efficient simulation of ultrasound fields and
pulse-echo responses is essential for transducer design, beamforming optimization,
and model-based imaging research. Full-wave methods, including finite-difference and
k-space pseudospectral approaches, provide accurate solutions of the acoustic wave
equations and can model complex propagation phenomena, but their computational cost
limits their use for large-scale simulations
\cite{cox_k_2007,pinton_heterogeneous_2009,bossy_three-dimensional_2004,
treeby_k-wave_2010}. In contrast, Spatial Impulse Response (SIR)-based methods provide
a computationally efficient alternative by expressing the acoustic field as the
superposition of elementary responses generated by the transducer aperture under
linear and homogeneous propagation assumptions. These methods are widely used for
ultrasound simulation, with Field II being one of the most established
implementations for time-domain pressure-field and pulse-echo calculations
\cite{jensen_jorgen_arendt_field_1996}.\\

The SIR formulation originates from the Tupholme-Stepanishen solutions for baffled
pistons \cite{tupholme_generation_1969,stepanishen_transient_1971,
stepanishen_time-dependent_1971}, which describe transient radiation as the
convolution between geometry-dependent impulse responses and excitation waveforms.
Jensen and Svendsen adapted this theory into a practical numerical framework by
subdividing the aperture into rectangular patches and summing their delayed
contributions at each observation point
\cite{jensen_calculation_1992}. This approach remains the basis of some ultrasound
simulators \cite{garcia_simus_2022,cigier_simus_2022}; however, the explicit sampling of each patch contribution becomes
increasingly expensive for large apertures, high sampling frequencies, dense field
evaluations, and pulse-echo simulations involving many transmit-receive interactions.\\

For rectangular apertures in the far field, the SIR of each patch has a trapezoidal
temporal shape. Conventional Fully Sampled Trapezoid (FST) implementations construct
this response by evaluating all temporal samples within the trapezoid support,
making the computational cost dependent on aperture dimensions and sampling
frequency. This dependency becomes particularly limiting for modern ultrasound
systems with large matrix arrays and high-resolution simulation requirements. \\

In this work, we introduce Sparse Delta Integration (SDI), a new formulation of the
far-field SIR that replaces the explicit trapezoidal representation by a sparse
distribution of Dirac delta functions. By exploiting the relationship between the
trapezoidal SIR and its second derivative, SDI represents each patch contribution
using only a fixed number of weighted temporal samples, followed by numerical
integration to recover the SIR. This reformulation removes the dependence on
trapezoid duration, enables efficient vectorized computation, and provides a
constant-cost alternative to FST accumulation.\\

Furthermore, we derive Spectral SDI, a frequency-domain extension of the method for
pulse-echo simulation. By expressing the SIR analytically in the Fourier domain,
Spectral SDI avoids redundant Fourier transforms required by conventional approaches,
reducing the computational cost of RF simulation while preserving the accuracy of
the original SIR formulation.\\

The proposed methods are implemented in eSDIva, an open-source Python framework for
ultrasound simulation supporting arbitrary transducer geometries, apodization,
electronic focusing, pressure-field evaluation, and pulse-echo RF computation.
Compared with Field II, eSDIva provides substantial acceleration while maintaining
high numerical agreement across a wide range of configurations. The proposed
framework enables large-scale, high-resolution ultrasound simulations and provides a
flexible foundation for future optimization, inverse problems, and computational
imaging applications.

\section{Sparse Delta Integration (SDI):}
\label{sec:SDI}

\subsection{The Spatial Impulse Response (SIR):}

The acoustic field generated by an ultrasound transducer can be described using the
Tupholme-Stepanishen formulation \cite{tupholme_generation_1969}
\cite{stepanishen_transient_1971}, which expresses the
field through the Spatial Impulse Response (SIR), briefly described in this section.
Consider an aperture of surface $S$ mounted in an infinite rigid baffle and vibrating with a
prescribed normal velocity $v_n(\mathbf{r},t)$. Assuming a homogeneous medium with sound
speed $c_0$, the resulting velocity potential at a field point $\mathbf{r}_p$ can be
expressed as

\begin{equation}
\phi (\mathbf{r_p},t)
=
v_n(t)
\overset{t}{*}
h(\mathbf{r_m},\mathbf{r_p},t)
\label{eq:vel_pot}
\end{equation}

where $\overset{t}{*}$ denotes convolution in time and $h(\mathbf{r_m},\mathbf{r_p},t)$
is the Spatial Impulse Response (SIR), defined as

\begin{equation}
h(\mathbf{r_m},\mathbf{r_p},t)
=
\frac{1}{2\pi}
\int_S
\frac{\delta\left(
t-\frac{|\mathbf{r_m}-\mathbf{r_p}|}{c_0}
\right)}
{|\mathbf{r_m}-\mathbf{r_p}|}
 dS.
\label{eq:SIR}
\end{equation}

The SIR represents the propagation operator between the transducer surface and the field
point. Consistent with Huygens' principle, it is obtained by summing the contributions
of all infinitesimal surface elements, each arriving at the field point after its
corresponding propagation delay. \\

From the velocity potential, the emitted acoustic pressure field is obtained as

\begin{equation}
    p_e(\mathbf{r_p},t) = \rho_0 \frac{\partial \phi (\mathbf{r_p},t)}{\partial t}
    \label{eq:press_field_and_vel_pot}
\end{equation}

Which yields to

\begin{equation}
    p_e(\mathbf{r_p},t) = \rho_0 \frac{\partial v_n}{\partial t}(t)
    \overset{t}{*} h(\mathbf{r}_{m,p},t) 
    \label{eq:press_field_sir}
\end{equation}

Thus, the emitted pressure field generated by a planar baffle \(m\) is obtained by
convolving the baffle normal velocity \(v_n(t)\), which encodes the temporal excitation,
with the SIR, which captures the spatial and geometric characteristics of the aperture.

\begin{figure*}[!b]
\centerline{\includegraphics[width=\textwidth]{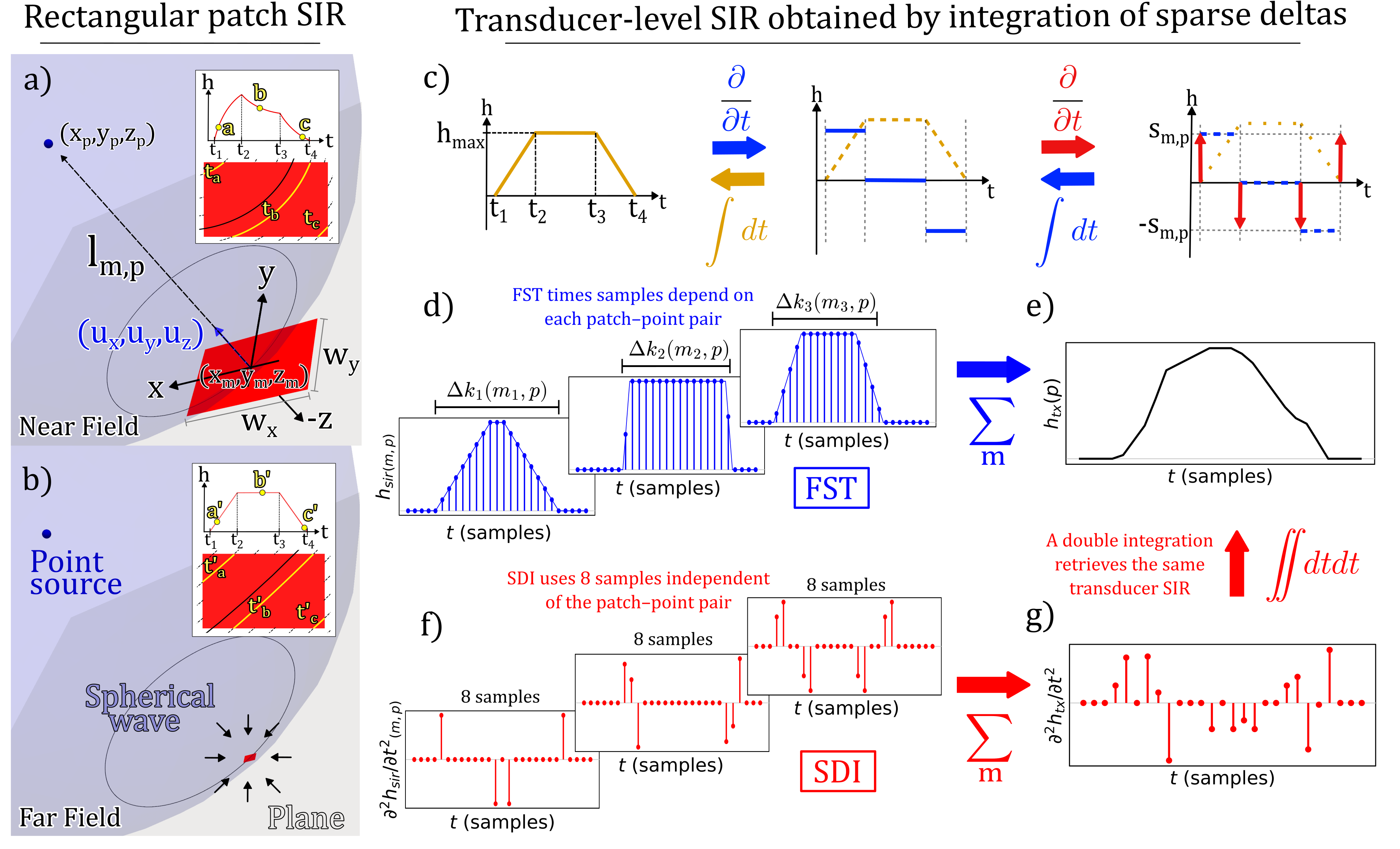}}

\caption{
\textbf{Trapezoidal far-field spatial impulse response (SIR) of a rectangular aperture
and the new Sparse Delta Integration (SDI) method.}
The SIR is obtained by acoustic spatial reciprocity as the intersection of a spherical
wave emitted from the field point with the aperture surface.
\textbf{(a)} Near-field configuration. A spherical wave emitted from the field point
intersects a rectangular aperture mounted on an infinite rigid baffle. The resulting SIR
and its 2D projection are shown in the upper-right corner. In the near field, curved
isodistance contours produce the characteristic curved SIR.
\textbf{(b)} Far-field configuration. The effective aperture decreases and the
isodistance contours become nearly planar, yielding a trapezoidal SIR.
\textbf{(c)} The trapezoidal SIR can be represented by four Dirac delta functions
through second-order differentiation and reconstructed by double integration, forming
the basis of the Sparse Delta Integration (SDI) method.
\textbf{(d)} Fully Sampled Trapezoid (FST), the discrete representation of the
trapezoidal SIR, where the number of occupied samples depends on the trapezoid width,
$\Delta k_i$, which varies for each patch--field-point pair $(m_i,p)$.
\textbf{(e)} Conventional computation of the transducer SIR, $h_{\mathrm{tx}}$, obtained
by summing the patch SIRs on the discrete time grid.
\textbf{(f)} Discrete SDI implementation. When the characteristic times
$(t_1,t_2,t_3,t_4)$ coincide with sampling instants, four discrete delta functions are
sufficient. Otherwise, each delta is linearly distributed between neighboring samples,
resulting in eight nonzero discrete deltas, independently of the patch or field point.
\textbf{(g)} Owing to the linearity of differentiation and summation, the discrete delta
contributions from all patches can be accumulated into a sparse delta distribution and
numerically integrated twice to recover the same transducer SIR as the conventional
approach shown in (e).}
\label{fig:rectang_sir}
\end{figure*}

\subsection{Fully Sampled Trapezoid (FST): Far-field SIR of a rectangular patch}

Using acoustic spatial reciprocity, the SIR formulation in (\ref{eq:SIR}) may be
interpreted by considering a spherical wave originating from the field point and
propagating back toward the aperture, rather than the aperture acting as an emitter.
The SIR is then reduced to determining the intersection between this
backward-projected wavefront and the aperture surface.  \\

For rectangular apertures, an exact solution to (\ref{eq:SIR}) was derived in
\cite{stepanishen_transient_1971, stepanishen_time-dependent_1971}. Geometrically, the
SIR at a given time corresponds to the fraction of the aperture lying on an isodistance
curve relative to the field point. In the near field, these curves form rounded
contours, whereas in the far field they can be approximated as straight lines as
illustrated in Fig.~\ref{fig:rectang_sir}(a)--(b). \\

In general, under the assumption of far-field, the SIR of a rectangular aperture tends
to a trapezoidal shape. Consequently, the trapezoidal SIR can be fully described by five
parameters: 

\begin{itemize}
\item $t_1$: time-of-flight \(TOF\) from the nearest aperture corner,
\item $t_2$,$t_3$: TOFs from the second and third nearest corners, which define the
    onset and end of the plateau,
\item $t_4$: TOF from the farthest corner,
\item $h_{max}$: the plateau height,
\end{itemize}

Placing the rectangular aperture $m$ of width and height of $w_{mx},w_{my}$ in the x-y
plane with its center at the coordinate $\mathbf{r_m}$, the field point position vector
- coming from the $m$-th rectangle to the $p$-th field point-is
$\mathbf{r}_{m,p}=\mathbf{r_p}-\mathbf{r_m}$, where $\mathbf{r_p}$ are the coordinates
of the field point.  This vector can be further decomposed into a distance
$l_{m,p}=||\mathbf{r}_{m,p} ||$ and an unit vector
$\mathbf{u}_{m,p}=\mathbf{r}_{m,p}/l_{m,p}=(u_x,u_y,u_z)$. Using this description, one
can get the following equations

\begin{align}
    t_1 &= l_{m,p}/c_0 -\frac{\Delta t_1+\Delta t_2}{2} \label{eq:TOFs1}\\
    t_2 &= t_1+\Delta t_1 \\
    t_3 &= t_1+\Delta t_2 \\
    t_4 &= t_1+\Delta t_1+ \Delta t_2
    \label{eq:TOFs}
\end{align}

Where $\Delta t_1$ and $\Delta t_2$, correspond to the min and max TOF for the side
lengths of the rectangle:

\begin{align}
    \Delta t_1 &= \min \left(\frac{w_{mx} |u_x|}{c_0},\frac{w_{my} |u_y|}{c_0}\right) \\
    \Delta t_2 &= \max \left(\frac{w_{mx} |u_x|}{c_0},\frac{w_{my} |u_y|}{c_0}\right)
    \label{eq:Deltat}
\end{align}

The shape of the trapezoid is a function of $\Delta t_1$ and $\Delta t_2$ and so is its amplitude, but the area 
of the trapezoid is equal to \cite{stepanishen_transient_1971}
\cite{jensen_calculation_1992}:

\begin{equation}
a_{rec} (l_{m,p})=\frac{w_{mx}w_{my}}{2 \pi l_{m,p}} 
    \label{eq:area_rec}
\end{equation}

Geometrically, the area can also be expressed as,

\begin{equation}
a_{rec}=h_{max} \Delta t_2 
\label{eq:area-hmax}
\end{equation}

Combining (\ref{eq:area_rec}) and (\ref{eq:area-hmax}) gives the trapezoid amplitude

\begin{equation}
h_{m,p}^{max}=\frac{w_{mx}  w_{my}}{2 \pi \Delta t_2 l_{m,p} }
\end{equation}

Thus, the far-field SIR of the \(m\)-th rectangular patch evaluated at the \(p\)-th field
point is represented by the Fully Sampled Trapezoid (FST) piecewise function

\begin{equation}
    h(\mathbf{r}_{m,p},t) =
\begin{cases}
s_{m,p} (t - t_1), & t_1 < t < t_2, \\
h^{\text{max}}_{m,p}, & t_2 < t < t_3, \\
s_{m,p} (t_4 - t), & t_3 < t < t_4, \\
0, & \text{elsewhere}.
\end{cases}
\label{eq:trapezoid_sir}
\end{equation}

With,

\begin{equation}
s_{m,p} = \frac{h^{\text{max}}_{m,p}}{\Delta t_1} =  \frac{h^{\text{max}}_{m,p}}{\Delta
t_2} .
\label{eq:slope_sir}
\end{equation}

\subsection{A new approach: The Sparse Delta Integration (SDI)}
\label{sec:SDI-method}

In this section we show how (\ref{eq:trapezoid_sir}) is equivalent to a
two temporal integrations of a sparse distribution of Dirac delta functions, the basis
of the Sparse Delta Integration (SDI) method. \\

As illustrated in Fig \ref{fig:rectang_sir} c, the first derivative FST
function in (\ref{eq:trapezoid_sir}) can be written as

\begin{equation}
\frac{\partial h(\mathbf{r}_{m,p},t)}{\partial t}
=
\begin{cases}
s_{m,p}, & t_1 < t < t_2, \\
- s_{m,p}, & t_3 < t < t_4, \\
0, & \text{elsewhere},
\end{cases}
\label{eq:first_deriv_sir_piecewise} 
\end{equation}

Introducing the differential operator ($D = \partial /\partial t$) and using the
Heaviside step function ($u(t)$) can be rewritten as

\begin{align}
    D h(\mathbf{r}_{m,p},t) &= s_{m,p} \sum_{i=1}^{4}
    \sigma_{i}  u(t-t_i) 
    \label{eq:first_deriv_sir_expansion}  \\
    \text{With, } (\sigma_{1}, \sigma_{2}, \sigma_{3}, \sigma_{4}) &=(1,-1,-1,1)
    \label{eq:sign}
\end{align}

Differentiating once more yields,

\begin{equation}
    D^2h(\mathbf{r}_{m,p}, t) = \Delta\delta(t)
    \label{eq:second_deriv_sir}
\end{equation}

With,

\begin{equation}
    \Delta\delta(t) = s_{m,p}\sum_{i = 1}^{4} \sigma_i \delta(t-t_i)
    \label{eq:4delta_comb}
\end{equation}

Where $\delta(t)$ denotes the Dirac delta distribution. Since $h(r_{m,p},t)=0$ for
$t<t_1$, the initial SIR piecewise function (\ref{eq:trapezoid_sir}) can be recovered
through two successive integrations

\begin{equation}
    h(\mathbf{r}_{m,p},t)
    = \int_\infty^{t} \int_\infty^{t'} \Delta\delta(t'') dt''dt'= I^2 \, \Delta\delta(t)
    \label{eq:double_int_sir}
\end{equation}

where $I$ denotes the integration operator $If=\int_{-\infty}^t f(t') dt'$. 
For compactness, subsequent derivations use
the operators $D$ and $I$ to represent differentiation and integration with respect to
time. Note that (\ref{eq:trapezoid_sir}) and (\ref{eq:double_int_sir}) describe
the same far-field trapezoidal SIR of a rectangular aperture. However,
(\ref{eq:double_int_sir}) is formulated from its second temporal derivative, represented
as a sparse distribution of Dirac delta functions followed by two temporal integrations.

\subsection{From rectangular patches to arbitrary transducer geometries}
\label{sec:conv_transducer_sir}

In the previous sections, the spatial impulse response (SIR) of a rectangular radiator
was derived under the far-field assumption. In practice, ultrasound transducers are
represented by subdividing each element into small rectangular patches, each satisfying
the far-field condition, while the total aperture response is obtained by summing the
patch contributions \cite{jensen_calculation_1992}. The patch dimensions must satisfy
\cite{kino_acoustic_1987}:

\begin{equation}
w \ll \sqrt{\frac{4lc_0}{f}},
\label{eq:far-field_cond}
\end{equation}

where $w$ is the largest patch dimension, $l$ is the distance to the field point, and
$f$ is the highest simulated frequency.

Apodization and electronic focusing are incorporated by weighting each patch with an
apodization coefficient $a_m$ and delaying its SIR by $\tau_m$. The transducer SIR is
therefore

\begin{equation}
h_{\mathrm{tx}}(\mathbf{r}_p,t)
\approx
\sum_{m=1}^{M} a_m\,h(\mathbf{r}_{m,p},t-\tau_m),
\label{eq:transducer_sir}
\end{equation}

where $h$ denotes the patch SIR, computed either by the FST formulation
(\ref{eq:trapezoid_sir}) or the SDI formulation (\ref{eq:double_int_sir}). Owing to the
linearity of integration, the SDI transducer SIR can be written as

\begin{equation}
h_{\mathrm{tx}}(\mathbf{r}_p,t)
\approx
I^2\!\left[
\sum_{m=1}^{M}
a_m\,\Delta\delta(t-\tau_m)
\right].
\label{eq:transducer_sir_SDI}
\end{equation}

\begin{figure*}[!b]
\centerline{\includegraphics[width=\textwidth]{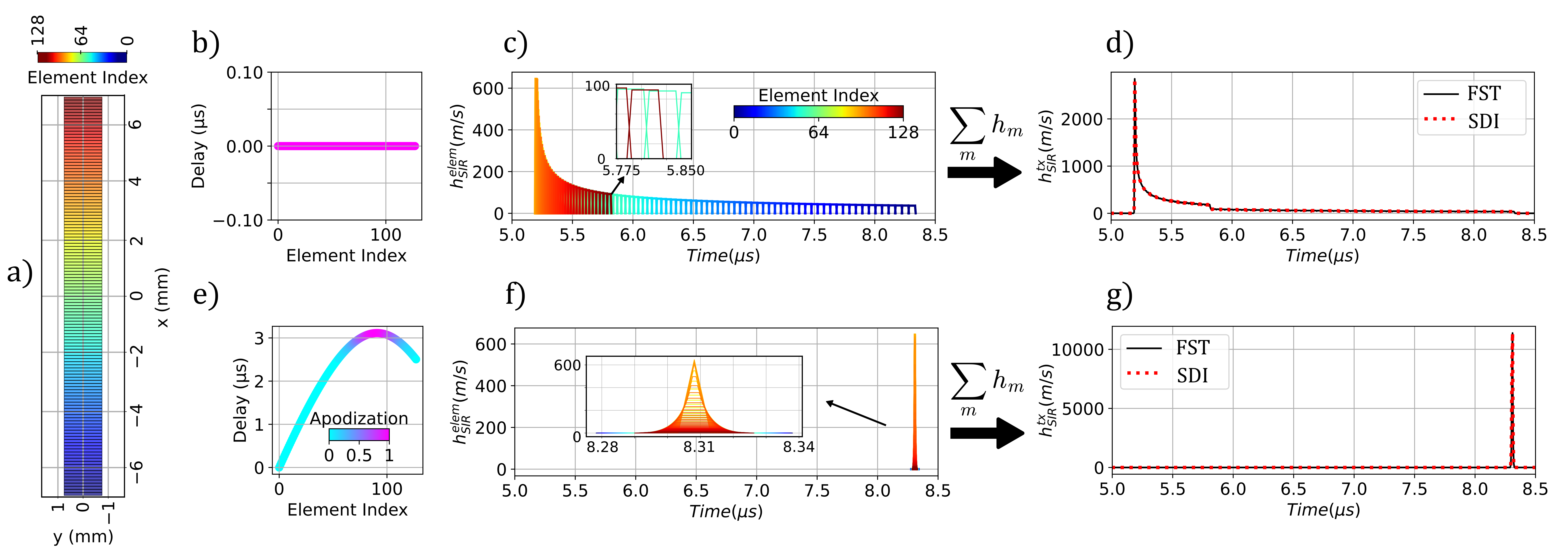}}
\caption{\textbf{Element-wise and accumulated transducer spatial impulse responses
    (SIRs) computed with the Fully Sampled Trapezoid (FST) and Sparse Delta Integration
    (SDI) methods for a linear array.}
\textbf{a)} Simplified geometry of a 128-element linear array without elevation
focusing or element subdivision, showing the transducer dimensions and element
numbering.
\textbf{b)} Element-wise propagation delays, color-coded by apodization weights, for
the case of zero delay law and uniform apodization.
\textbf{c)} Trapezoidal SIR of each element, color-coded by element index.
\textbf{d)} The corresponding transducer SIR obtained by summing the element SIRs on the
discrete time grid.
\textbf{e)--g)} Same as b)--d), but using a delay law focused at $(4,0,8)$~mm and a
Hamming apodization with $F/D = 1$. The accumulated transducer SIR computed with the
conventional approach (FST) and the SDI method is identical.}

\label{fig:tx_sir}
\end{figure*}

\subsection{Discrete Time Implementation of FST and SDI} \label{sec:discrete_implementation}

So far, the derivation of (\ref{eq:transducer_sir}) and
(\ref{eq:transducer_sir_SDI}) has been expressed in continuous time. For numerical
implementation, however, the temporal axis must be discretized. The continuous variable
\(t\) is replaced by its sampled counterpart \(\{t_k\}\), consisting of \(T\) samples
with sampling interval \(\Delta t = 1/f_s\) (see
Appendix~\ref{appendix:discretization}). \\

For each field point \(p\), the algorithm loops over all patches
\(m \in \{1,\ldots,M\}\) and computes the trapezoidal parameters
\((t_1,t_2,t_3,t_4,h_{\max})\). These define the discrete index bounds

\begin{equation}
k_{1}^{(m,p)}
=
\left\lfloor
\frac{t_1-t_0}{\Delta t}
\right\rfloor,
\qquad
k_{4}^{(m,p)}
=
\left\lceil
\frac{t_4-t_0}{\Delta t}
\right\rceil,
\label{eq:discrete_index_bounds}
\end{equation}

where \(\lfloor \cdot \rfloor\) and \(\lceil \cdot \rceil\) denote the floor and
ceiling operators, respectively. The corresponding trapezoid span in samples is

\begin{equation}
\Delta k_{(m,p)}
=
k_{4}^{(m,p)} - k_{1}^{(m,p)}.
\label{eq:deltak}
\end{equation}

The SIR contribution of patch \(m\) to field point \(p\) can then be computed using
either the FST or SDI formulation.

\subsubsection*{FST method} \label{sec:naive_method}

When the Fully Sampled Trapezoid (FST) method is selected, for each field point-patch pair $(p,m)$ the trapezoid is explicitly evaluated at 
every sample $k \in [k_{\mathrm{start}}, k_{\mathrm{end}}]$ using the piecewise expression in (\ref{eq:trapezoid_sir}). 
The per-patch apodization and delay are applied, and contributions are accumulated as

\begin{equation}
    h_{\mathrm{out}}[p,k] = \sum_m a_m  h \left(\mathbf{r}_{m,p}, t_k - \tau_m\right),
\label{eq:naive_accumulation}
\end{equation}

Repeating this process over all patches yields the final 
tensor $h_{\mathrm{out}} \in \mathbb{R}^{P \times T}$, containing the SIR at every field
point.\\

\subsubsection*{SDI method} \label{sec:SDI_method}

The Sparse Delta Integration (SDI) method reformulates the trapezoidal SIR using delta functions located at 
$\{t_1,t_2,t_3,t_4\}$. Since these times do not generally coincide with the discrete temporal grid, each impulse 
is conservatively split between the nearest floor and ceiling indices using linear interpolation weights 
as depicted in Fig. \ref{fig:rectang_sir} d. This preserves the total contribution while remaining consistent with the discrete 
representation. As a result, the original four impulses are mapped to eight discrete
contributions. \\

For a continuous time $t_i$, the fractional grid index is

\begin{equation}
    k_{i} = k(t_i) = \frac{t_i - t_0}{\Delta t} \in \mathbb{R},
\label{eq:fractional_index}
\end{equation}

with interpolation weights $w_{\mathrm{ceil, floor}}(t_i) \in [0,1]$ defined as

\begin{align}
    w_{\mathrm{ceil}}(t_i) = w_{c}(t_i)  &= k_i - \lfloor k_i \rfloor \\
    w_{\mathrm{floor}}(t_i) = w_{f}(t_i) &= \lceil k_i \rceil - k_i = 1 - w_{c}(t_i)
\label{eq:interpolation_weights}
\end{align}

For example, when ($k_i=\lfloor k_i\rfloor$), then ($w_{\mathrm{ceil}}=0$), meaning that
the interpolated value is given entirely by the floor sample. \\

Let $t^{f}_{i}$ and $t^{c}_{i}$ denote the times discrete times on the temporal grid
(\ref{eq:temporal_grid}) at the floor and ceiling indices of $k_i$, respectively. For
example, for a continuous time $t_i$,

\begin{align}
    t^f_{i} &= t^f_{k_i} = t_{\lfloor k_i \rfloor}\\
    t^c_{i} &= t^c_{k_i} = t_{\lceil k_i \rceil}
    \label{eq:ceil_floor_times}
\end{align}

Thus, the discrete approximation of the continuous second-derivative impulses, 
denoted $\Delta\delta(t_k)$, is decomposed as

\begin{equation}
\Delta\delta(t_k) = \Delta\delta^{f}(t_k) + \Delta\delta^{c}(t_k),
\label{eq:delta_decomposition}
\end{equation}

with floor deltas distributions, $\Delta \delta^{f}(t)$, defined as

\begin{align}
    \Delta\delta^{f}(t_k) &=s_{m,p}\sum_{i=1}^4 \sigma_i
    w_{f}(t_i)\delta(t_k - t_i^{f}), 
    \label{eq:delta_floor_ceil} 
\end{align}

and ceil deltas distribution $\Delta \delta^{c}(t)$ defined by the same equation
(\ref{eq:delta_floor_ceil}) but with ceil terms. \\

These weighted impulses are accumulated into an intermediate array $\mathrm{d2h}$ as

\begin{equation}
\mathrm{d2h}[p,k] = \sum_m a_m  \Delta\delta(t_k - \tau_m).
\label{eq:SDI_accumulation}
\end{equation}

If any SDI events are accumulated for a given field point, the trapezoidal SIR is recovered by performing two discrete 
integrations along the temporal axis of $\mathrm{d2h}[p,:]$. The result is then added to
$h_{\mathrm{out}}[p,:]$. Both integrations are implemented as 
cumulative sums along the time axis. 

%
\subsection{Computational Cost of FST and SDI} \label{sec:cost_analysis}

The two methods differ only in how the trapezoidal SIR is accumulated over the
temporal grid. FST evaluates each patch's trapezoid (\ref{eq:trapezoid_sir}) explicitly
over its $\Delta k$-sample support (\ref{eq:deltak}); writing $\overline{\Delta k}$ for the
mean width, the cost per field point is $M\,\overline{\Delta k}$. SDI instead writes eight
Dirac deltas per patch and recovers the SIR with two integrations over the $T$-sample
grid, costing $8M + 2T$. SDI is therefore cheaper when

\begin{equation}
    \overline{\Delta k} \gg 8 + \frac{2T}{M},
\label{eq:heuristic_condition}
\end{equation}

i.e. when each trapezoid spans more than the eight samples SDI spends per patch plus the
integration cost amortised over the $M$ patches. Expressed geometrically
(Appendix~\ref{appendix:heuristic_condition}) through the aperture extent $A$, patch width
$w$, and maximal propagation distance $l_{\max}$, this becomes

\begin{equation}
    A \gg 2\,\frac{1+\varepsilon}{1 - w_{crit}/w}\,w\,l_{\max},
\label{eq:final_heuristic_condition}
\end{equation}

valid for patches above the critical size

\begin{equation}
    w > w_{crit} = \frac{8 c_0}{f_s}.
\label{eq:critical_patch_size}
\end{equation}

In practice $f_s \sim 10 f_c$, yielding

\begin{equation}
    w_{crit} \sim \lambda_0
    \label{eq:patch_bigger}
\end{equation}

Here, $\varepsilon$ is a small field 
distances correction (~$l_{\min}/l_{\max}$) that vanishes for large fields. The critical
size $w_{crit}$ is the patch width below which a trapezoid spans fewer than eight 
samples, so SDI's 
fixed eight-delta cost never amortises. Above $w_{crit}\sim \lambda_0 $, SDI grows more 
favourable as the aperture enlarges since more contributing patches mean more FST 
trapezoid evaluations, whereas for sub-critical patches or low sampling frequencies 
($w < w_{crit}$) FST might outperform SDI.

\subsection{From SIR to Emitted Pressure Field:}
 \label{sec:emission}

Here, we describe how to derive acoustic pressure fields from time-domain spatial
impulse responses (SIRs), a computation directly accelerated by SDI. \\ 

For a pulsed excitation defined as \( v_n(t)=\delta(t) \), the emitted pressure field
\( p_{e,\delta} \) follows directly from
(\ref{eq:press_field_and_vel_pot}), yielding

\begin{equation}
    p_{e,\delta}(\mathbf{r},t) \propto h(\mathbf{r},t)
\label{eq:press_field_proportionality}
\end{equation}

%
%
%
A similar proportional relationship as (\ref{eq:press_field_proportionality}) can be
derived for the monochromatic pressure field by evaluating
(\ref{eq:press_field_and_vel_pot}) in the frequency domain 
at the transducer central angular frequency \( \omega_c = 2\pi f_c \), yielding

\begin{equation}
    p_{e,\omega=\omega_c}(\mathbf{r}_p)
    \propto
    \left| H(\mathbf{r}_p,\omega = \omega_c) \right|
\label{eq:monochromatic_pressure_field}
\end{equation}

Thus, in a homogeneous and lossless medium, the spatial distribution of the
monochromatic pressure field is directly determined by the magnitude of the Fourier
transform of the SIR evaluated at \( \omega_c \). \\

For arbitrary excitations \( v_n(t) \), the pressure field
(\ref{eq:press_field_and_vel_pot}) is formulated in the
frequency domain where convolution becomes multiplication. The resulting pressure field
is then obtained by evaluating the product of the Fourier transforms of the excitation,
the SIR, and any additional transfer functions accounting for frequency-dependent
effects such as transducer bandwidth, attenuation, and dispersion, as proposed in
\cite{jensen_ultrasound_1993}.  

%
%

\begin{figure*}[!b]
\centerline{\includegraphics[width=\textwidth]{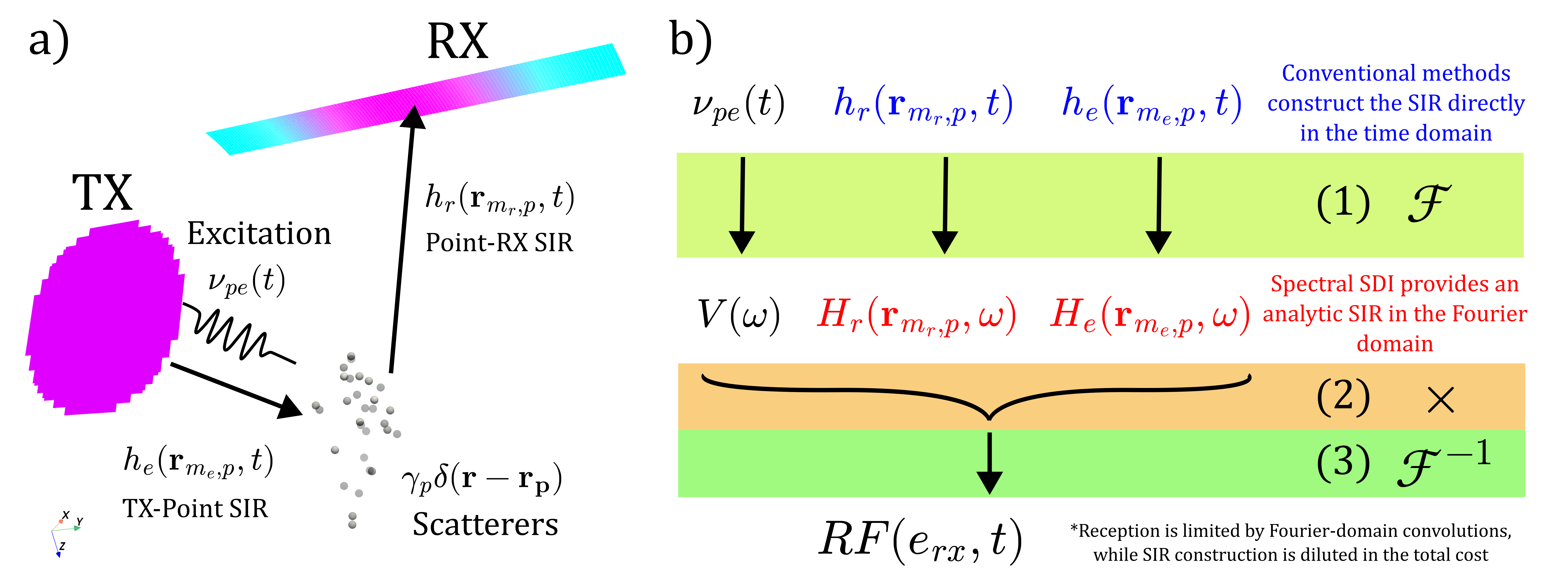}}

\caption{\textbf{Pulse-echo signal formation and Fourier-domain implementation of the
    Sparse Delta Integration (SDI) method.}
\textbf{a)} Schematic of pulse-echo signal formation. The emission aperture (TX),
driven by a defined excitation waveform, insonifies a distribution of scatterers
located at $\mathbf{r}_p$ with backscattering amplitudes $\gamma_p$, while the receive
aperture (RX) records the echoes. The pulse-echo spatial impulse response, $h_{pe}$, is
obtained by convolving the emission and reception SIRs, $h_e$ and $h_r$. The received RF
signal is then computed by convolving $h_{pe}$ with the pulse-echo velocity waveform,
$\nu_{pe}$, and summing the contributions from all scatterers and receive-aperture
patches.
\textbf{b)} Conventional implementation computes $h_e$ and $h_r$ in the time domain
using either the FST or SDI representation, followed by
Fourier-domain convolutions between $h_e$, $h_r$, and $\nu_{pe}$. This requires three
forward Fourier $\mathcal{F}$ transforms (for $h_e$, $h_r$, and $\nu_{pe}$) and one
inverse Fourier $\mathcal{F}^{-1}$ transform. In contrast, the proposed SDI formulation
provides a closed-form expression for the SIRs directly in the Fourier domain,
eliminating two forward Fourier transforms per field point.
}

\label{fig:pulse_echo_SDI}
\end{figure*}

\subsection{From SIR to Received Signals and SDI variants in Reception:} 
\label{sec:reception}

\subsubsection{Conventional method for simulation of RF data:}
Under the Born approximation - valid for weakly scattering media where multiple
scattering is neglected \cite{jensen_model_1991} - the received pressure at
position $\mathbf{r_m}$ is expressed as a surface integral of the scattered pressure
field over the aperture $S$:

\begin{equation}
    p_r(\mathbf{r_{m}}, t) = E(t) \overset{t}{*} \int_S p_s(\mathbf{r_{p,m}}, t)\, dS
\label{eq:received_pr_origin}
\end{equation}

where \(E_e(t)\) denotes the electromechanical impulse response of the receiving
elements. Using the scattering formalism introduced by
\cite{angelsen_theoretical_1980} and \cite{jensen_model_1991}, (\ref{eq:received_pr_origin})
can be rewritten compactly as

\begin{equation}
    p_r(\mathbf{r_{m_r}, t}) = v_{pe}(t) \overset{t}{*} f_m(\mathbf{r_p})
    \overset{\mathbf{r}}{*} h_{pe}(\mathbf{r}_{m_r,p}, t)
    \label{eq:received_pr}
\end{equation}

With,

\begin{align}
    f_m(\mathbf{r_p}) &= \frac{\Delta\rho(\mathbf{r_p})}{\rho_0} - 2 \frac{\Delta
    c(\mathbf{r_p})}{c_0}
    \label{eq:f_perturbations} \\
    h_{pe}(\mathbf{r},t) &= h^e(\mathbf{r}_{p,m_r},t) \overset{t}{*}
    h^r(\mathbf{r}_{m,p},t)
    \label{eq:pe_hsir} \\
    v_{pe}(t) &= \frac{\rho_0}{2c_0^2} E_e(t) \overset{t}{*} D^3
    v(t)
    \label{eq:pe_velocity}
\end{align}

Here, \(f_m\) is the medium scattering function, which encodes local density and
sound-speed perturbations; \(h_{pe}\) is the modified pulse-echo spatial impulse
response (SIR), relating the transducer geometry to the spatial distribution of the
scattered field; and \(v_{pe}\) is the pulse-echo velocity waveform incorporating the
transducer excitation and the electromechanical impulse responses during emission and
reception. Superindex with $r$ and $e$ and $m_r$ and $m_e$ denote the terms 
and the patches associated with the apertures used for reception and emission, RX and
EX, respectively. \\

In practice, the temporal derivatives and electromechanical responses are accounted for
by the transducer impulse responses, ($IR^e =h^e(t)$) and ($IR^r=h^r(t)$), -- 
not be confused with the transducer spatial impulse response (SIR).

\begin{equation}
    v_{pe} \propto \nu_{pe}(t) = e(t) \overset{t}{*} IR_{e/r}(t) \overset{t}{*} IR_{r/e}(t)
    \label{eq:v_pe}
\end{equation}

The final RF signal received in each corresponds to the sum of each scatterers signal 
received over the patches of the each RX apertures element. Then, if $E_{tx}$ are the
patches per element, and each scatterer is modelled as $f_m(\mathbf{r}_p) =
\gamma_p\delta(\mathbf{r}-\mathbf{r}_p)$ with $\gamma_p$ the scattering amplitude. The received
signal at the $e_{rx}$-th RX element is given by:

\begin{equation}
    RF(e_{rx}, t) = \sum_{m_e=1}^{M_e} \sum_{m_e=1}^{E_{rx}} \sum_{p=1}^{P} \gamma_p 
    \left[ \nu_{pe}(t) \overset{t}{*} h_{pe}(\mathbf{r}_{m_e,p}, \mathbf{r}_{p,m_r},t)
    \right]
\label{eq:RF_signal}
\end{equation}

Which can be developped for the independent SIRs of the emission and reception apertures
as:

\begin{multline}
    RF(e_{rx}, t) = \nu_{pe}(t) \overset{t}{*} \\ \left[ \sum_{p=1}^{P}
        \gamma_p \left [ h_{tx}(\mathbf{r}_{m_e,p},t) \overset{t}{*}
    h_{E_{rx}}(\mathbf{r}_{p,m_r},t) \right] \right]
\label{eq:RF_conventional_signal}
\end{multline}

Equation (\ref{eq:RF_conventional_signal}) defines the conventional formulation for RF
signal computation, where the pulse-echo response is obtained by explicit convolution of
the emission and reception SIRs, each evaluated using either the FST or SDI
representation. 

\subsubsection{The new Spectral SDI for RF data simulation}

Starting from (\ref{eq:pe_hsir}), the SDI formulation expresses the pulse-echo SIR in
terms of the second derivatives of the transmit and receive one-way SIRs:

\begin{align}
h_{pe}(t)
&= I^4 \big[ \Delta\delta^{e}(\mathbf{r}_{m_e,p},t)
\overset{t}{*}
\Delta\delta^{r}(\mathbf{r}_{p,m_r},t) ,\big]
\label{eq:conv_deltas_reception_text} \\
&= I^4\Delta\delta_{pe}(\mathbf{r}_{m_e,p},\mathbf{r}_{p,m_r},t),
\label{eq:h_pe_SDI_text}
\end{align}

where $\Delta\delta_{pe}$ is the pulse-echo delta distribution resulting from the
convolution of the transmit and receive delta distributions and contains $16$
$(4\times4)$ Dirac delta terms (Appendix~\ref{appendix:SDI_pulse_echo}). \\

Equations~(\ref{eq:conv_deltas_reception_text})--(\ref{eq:h_pe_SDI_text}) lead to two
equivalent computational formulations, depending on how the convolution in
(\ref{eq:RF_signal}) is evaluated. In the time domain, direct convolution with
(\ref{eq:h_pe_SDI_text}) yields the \emph{paired SDI} formulation introduced in Appendix
\ref{appendix:SDI_pulse_echo}.
Alternatively, in the Fourier domain, convolution with
(\ref{eq:conv_deltas_reception_text}) becomes multiplication and the
integration operators $I^n$ become $(j\omega)^{-n}$ scaling factors, yielding the
\emph{spectral SDI} formulation explained below. Their derivations are
provided in Appendix~\ref{appendix:SDI_pulse_echo}; only the final expression for
spectral SDI is presented here. \\

Equation (\ref{eq:RF_signal}) can
be evaluated with (\ref{eq:conv_deltas_reception_text}) in the Fourier
domain. Applying the convolution theorem the resulting formulation is

\begin{equation}
RF(e_{rx},t) =
\mathcal{F}^{-1}
\Biggl\{
\frac{V(\omega)}{(j\omega)^4}
\sum_{p=1}^{P} \gamma_p
\Bigl[
\Sigma_{TX} \times \Sigma_{RX}
\Bigr]
\Biggr\}_{(t)}
\label{eq:RF_spectral_SDI}
\end{equation}

with

\begin{multline}
    \Sigma_{TX/RX} = \sum_{m_{e/r}=1}^{M_e/E_{rx}} \Bigl[ s_{p,m_{e/r}}
\sin\left(\frac{\omega \Delta t_1^{e/r}}{2}\right) \\
\sin\left(\frac{\omega \Delta t_2^{e/r}}{2}\right)
e^{-j\omega t_1^{e/r}} \Bigr]
\label{eq:spectral_deltas}
\end{multline}

The "$/$" does the separation of the terms to be used in emission ($TX, M_e, m_{e},
t_i^e$) and reception ($RX, E_{rx}, m_{r}, t_i^r)$, respectively.
Unlike the conventional formulation, which explicitly constructs and samples the
transmit and receive temporal SIRs before convolution, the spectral SDI form directly 
evaluates their analytic spectra. The transmit and receive sums remain separable, so the
computation scales with the sum of the transmit and receive patch counts rather than
their product ($M_{tx}+E_{rx}\sim \mathcal{O}(M)$), and avoids the step of the forward
fourier transform of each SIR. \\

The paired and spectral SDI in reception formulations are mathematically identical and
differ only in the domain in which the integration and convolution are evaluated. The
paired form operates through time-domain placement of pulse-echo corner events, whereas
the spectral form exploits the separability of transmit and receive contributions in the
Fourier domain. 

\section{Methods}\label{sec:methods}

The FST and SDI methods were implemented in an open-source Python framework named
eSDIva, available at \url{}. eSDIva provides an implementation of the methods
presented in this work for computing Spatial Impulse Responses (SIRs), as
well as the corresponding emission and reception routines. The framework follows a modular
design, with dedicated components for transducer definition, simulation setup,
beamforming and post-processing, and visualization.  \\

Given the underlying formulation of the widely used reference tool Field II
\cite{jensen_jorgen_arendt_field_1996}, the FST method is expected to be conceptually aligned with
Field II’s approach to SIR computation, which also relies on explicit trapezoidal
integration. For this reason, Field II constitutes the appropriate reference for
assessing both the accuracy and computational performance of the proposed FST and SDI
methods, and for benchmarking their implementation within modern Python‑based tools. \\

All simulations were performed on a desktop workstation equipped with an AMD Ryzen 7
8700F processor (16 threads, up to 4.1 GHz) and 128GB of RAM, running Windows 11 Pro
(64-bit). Field II simulations were executed in MATLAB, whereas eSDIva used
Numba-accelerated Python code with just-in-time (JIT) compilation. No GPU acceleration
was employed; all computations were performed on the CPU to ensure a consistent
comparison across methods.

\subsection{Linear and Matrix Array SIR and Emitted Pressure:}

\subsubsection{SIR Computation Time and Accuracy}

The Spatial Impulse Response (SIR) was evaluated
for the linear and matrix array transducers described in the previous section using the
SDI and FST methods, as well as Field II's \texttt{calc\_h} function as a reference
implementation. \\

For this application, the spatial domain was defined as $x,y \in [-2,2]\mathrm{mm}, z
\in [3,13]\mathrm{mm}$. The number of spatial samples was varied as $n_{xyz} = n_x = n_y
= n_z \in \{9, 41, 81\}$, resulting in a total number of field points $P = n_{xyz}^3$.
Three sampling frequencies,$f_s \in \{100, 200, 300\}\mathrm{MHz}$, were tested to vary
both the total number of temporal samples $T$ and the number of samples per trapezoidal
event $\Delta k$.\\

Two main transducer geometries where used for the
benchmarking:

\begin{itemize}
\item \textbf{The\ linear\ array\ (12.5\ MHz):} consisted of 128 elements with an element height of 1.5 mm, width of 
108 $\mu$m, and $(n_x,n_y)=(1,10)$ subdivisions per element and elevation focus at 8mm.
\item \textbf{The\ matrix\ array\ (10MHz):} consisted of 55x55 elements with an element width and 
height of 290 $\mu$m, a pitch of 300 $\mu$m in both directions, and $(n_x,n_y)=(2,2)$ subdivisions per element.
\end{itemize}

For both linear and matrix arrays not apodization was used and delays where set to
focalize at $(0,0,8)mm$, thus a creating a converging wave (CW) setting. 
In addition, the subdivision along each dimension $n_x,n_y$ was
varied to modify the patch dimensions $w_x$, $w_y$, and consequently the total number of
patches $M$, as listed in Table \ref{tab:tableI}. \\ 

\begin{table}[h]

\newcommand{\cdarkgrey}{\cellcolor[rgb]{.851,.851,.851}}
\newcommand{\clightgrey}{\cellcolor[rgb]{.949,.949,.949}}

\centering\caption{Simulation parameter ranges for linear and matrix array transducers.}
\label{tab:tableI}

\begin{tabular}{cccc}
\cmidrule{1-4}
\multicolumn{2}{c|}{\cdarkgrey\textbf{Linear}} &
\multicolumn{2}{c}{\cdarkgrey\textbf{Matrix}} \\

\cmidrule{1-4}
\clightgrey\textbf{($n_x,n_y$)} &
\clightgrey\textbf{M} &
\clightgrey\textbf{($n_x,n_y$)} &
\clightgrey\textbf{M} \\

\cmidrule{1-4}
(1,10) & 1280 & (1,1) & 3025  \\
(2,20) & 5120 & (2,2) & 12100 \\
       &      & (3,3) & 27225 \\

\cmidrule{1-4}
\clightgrey\textbf{$f_s$ (MHz)} &
\clightgrey\textbf{T} &
\clightgrey\textbf{$f_s$ (MHz)} &
\clightgrey\textbf{T} \\

\cmidrule{1-4}
100 & 1034 & 100 & 1485 \\
200 & 2067 & 200 & 2970 \\
300 & 3100 & 300 & 4455 \\

\cmidrule{1-4}
\clightgrey\textbf{$n_{xyz}$} &
\clightgrey\textbf{P} &
\clightgrey\textbf{$n_{xyz}$} &
\clightgrey\textbf{P} \\

\cmidrule{1-4}
9 & 729 & 9 & 729 \\
41 & 68921 & 41 & 68921 \\
81 & 531441 & 81 & 531441 \\

\cmidrule{1-4}
\end{tabular}

\smallskip
\footnotesize
P: number of field points, T: number of time samples, M: number of patches

\end{table}

The SIR computation time was recorded for each method and configuration, and the
relative computation time ($CT$) of the SDI method with respect to each method 
was computed using the following expression:

\begin{equation}
\mathrm{CT} = 
\frac{\mathrm{Time}_{\mathrm{SDI}}}{\mathrm{Time}_{\mathrm{(FST,Field II)}}},
\label{eq:speedup}
\end{equation}

The mean squared error (MSE) between both SIRs solutions was computed as

\begin{equation}
    \mathrm{MSE}
    =
    \frac{1}{PT}
    \sum_{p=1}^{P} \sum_{k=1}^{T}
    \left|
    h_{\mathrm{eSDIva}}(\mathbf{r}_p,t_k)
    -
    h_{\mathrm{Field II}}(\mathbf{r}_p,t_k)
    \right|^2.
    \label{eq:mse}
\end{equation}

The computation times and MSE computed for these configurations where stored in Table
\ref{tab:tableII}. \\ 

\begin{table*}[!t]

\newcommand{\cdarkgrey}{\cellcolor[rgb]{.851,.851,.851}}
\newcommand{\clightgrey}{\cellcolor[rgb]{.949,.949,.949}}
\newcommand{\cred}{\cellcolor[rgb]{ .961,  .71,  .71}}
\newcommand{\corange}{\cellcolor[rgb]{ .961,  .875,  .663}}
\newcommand{\cgreen}{\cellcolor[rgb]{ .855,  .949,  .816}}
\newcommand{\cyellow}{\cellcolor[rgb]{ .988,  .992,  .773}}

\centering\caption[Comparison of SIR computation time and accuracy of eSDIva methods
relative to Field II]{Comparison of SIR computation time and accuracy for linear and
    matrix transducers using Field II and eSDIva (FST, SDI) under varying grid and
subdivision settings.}
\label{tab:tableII}

\begin{tabular}{w{c}{2.785em} w{c}{2.785em} w{c}{2.785em} w{c}{2.785em} w{c}{4em} w{c}{6em} w{c}{3.5em} w{c}{3.5em} w{c}{4em} w{c}{4em} c}

\cmidrule{7-10}
\multicolumn{6}{c}{} &
\multicolumn{2}{c|}{\cdarkgrey\textbf{Method time/SDI time}} &
\multicolumn{2}{c}{\cdarkgrey\textbf{MSE (\%)}} &
\\

\cmidrule{1-10}
\multicolumn{1}{w{c}{2.785em}}{\cdarkgrey\textbf{P}} &
\multicolumn{1}{w{c}{2.785em}}{\cdarkgrey\textbf{T}} &
\multicolumn{1}{w{c}{2.785em}}{\cdarkgrey\textbf{M}} &
\multicolumn{1}{w{c}{2.785em}}{\cdarkgrey\textbf{$\Delta k$}} &
\multicolumn{1}{w{c}{4em}}{\cdarkgrey\textbf{8+2T/M}} &
\multicolumn{1}{w{c}{6em}|}{\cdarkgrey\textbf{SDI time (s)}} &
\cdarkgrey\textbf{FST} &
\multicolumn{1}{c|}{\cdarkgrey\textbf{Field II}} &
\cdarkgrey\textbf{FST} &
\cdarkgrey\textbf{SDI} &
\\

\cmidrule{1-11}
\clightgrey 1331 & 725 & 1280 & 6.45 & 9.13 & \cgreen 0.005 $\pm$ 0.001 &
\corange 1.07 & \cred 26.63 &
0.00004 & 0.00004 &
\multirow{8}{*}{%
  \rotatebox{90}{\textbf{Linear Array}}%
}
\\

\clightgrey 68921 & 725 & 1280 & 6.34 & 9.13 & \cgreen 0.220 $\pm$ 0.006 &
\corange 1.27 & \cred 31.38 &
0.00002 & 0.00002 & \\

\clightgrey 531441 & 725 & 1280 & 6.33 & 9.13 & \cgreen 1.757 $\pm$ 0.020 &
\corange 1.20 & \cred 30.25 &
0.00002 & 0.00002 & \\

\cmidrule{1-10}
68921 & \clightgrey 725 & 1280 & 6.34 & 9.13 & \cgreen 0.220 $\pm$ 0.006 &
\corange 1.27 & \cred 31.38 &
0.00002 & 0.00002 & \\

68921 & \clightgrey 1444 & 1280 & 10.22 & 10.26 & \cgreen 0.273 $\pm$ 0.005 &
\corange 1.42 & \cred 30.31 &
0.00002 & 0.00002 & \\

68921 & \clightgrey 2164 & 1280 & 14.20 & 11.38 & \cgreen 0.324 $\pm$ 0.003 &
\corange 1.54 & \cred 28.86 &
0.00002 & 0.00002 & \\

\cmidrule{1-10}
68921 & 725 & \clightgrey 1280 & 6.34 & 9.13 & \cgreen 0.220 $\pm$ 0.006 &
\corange 1.27 & \cred 31.38 &
0.00002 & 0.00002 & \\

68921 & 725 & \clightgrey 5120 & 4.64 & 8.28 & \cgreen 0.712 $\pm$ 0.004 &
\corange 1.12 & \cred 20.40 &
0.00008 & 0.00008 & \\

\cmidrule{1-11}
\rowcolor[rgb]{.851,.851,.851}
 &  &  &  & & & & & & & \\

\cmidrule{1-11}
\clightgrey 1331 & 767 & 3025 & 17.11 & 8.51 & \cgreen 0.012 $\pm$ 0.003 &
\corange 2.32 & \cred 139.32 &
0.00003 & 0.00003 &
\multirow{9}{*}{%
  \rotatebox{90}{\textbf{Matrix Array}}%
}

\\

\clightgrey 68921 & 767 & 3025 & 17.04 & 8.51 & \cgreen 0.469 $\pm$ 0.005 &
\corange 2.44 & \cred 186.46 &
0.00003 & 0.00003 & \\

\clightgrey 531441 & 767 & 3025 & 17.03 & 8.51 & \cgreen 3.666 $\pm$ 0.027 &
\corange 2.46 & \cred 186.72 &
0.00003 & 0.00003 & \\

\cmidrule{1-10}
68921 & \clightgrey 767 & 3025 & 17.04 & 8.51 & \cgreen 0.469 $\pm$ 0.005 &
\corange 2.44 & \cred 186.46 &
0.00003 & 0.00003 & \\

68921 & \clightgrey 1525 & 3025 & 31.99 & 9.01 & \cgreen 0.556 $\pm$ 0.007 &
\corange 3.48 & \cred 179.65 &
0.00001 & 0.00001 & \\

68921 & \clightgrey 2287 & 3025 & 46.96 & 9.51 & \cgreen 0.675 $\pm$ 0.010 &
\corange 3.91 & \cred 165.47 &
0.00001 & 0.00001 & \\
\cmidrule{1-10}
68921 & 767 & \clightgrey 3025 & 17.04 & 8.51 & \cgreen 0.469 $\pm$ 0.005 &
\corange 2.44 & \cred 186.46 &
0.00003 & 0.00003 & \\

68921 & 767 &\clightgrey 12100 & 9.61 & 8.13 & \cgreen 1.659 $\pm$ 0.021 &
\corange 1.78 & \cred 70.87 &
0.00006 & 0.00006 & \\

68921 & 767 & \clightgrey 27225 & 7.18 & 8.06 & \cgreen 3.659 $\pm$ 0.034 &
\corange 1.50 & \cred 42.40 &
0.00004 & 0.00004 & \\

\cmidrule{1-11}

\end{tabular}

\smallskip
\footnotesize
P: number of field points, T: number of time samples, M: number of patches,
$\Delta k$: average trapezoid width in samples.
Columns FST and Field II under \emph{Method time/SDI time} give each method's SIR
wall time normalized by the SDI time (green column); a ratio $>1$ means slower than SDI.
For instance, for the matrix transducer at $P=531,441$ Field II takes
$186.72\times3.67\approx685\,\mathrm{s}\approx11\,\mathrm{min}$ against SDI's
$3.67\,\mathrm{s}$. Yellow, orange and red mark the fastest, intermediate and slowest of
these three columns; green marks a method faster than SDI (ratio $<1$).

\end{table*}

\subsubsection{Emitted Pressure Visualization}

For visual assessment, the SIR $h_{sir}$ of size $[P,T]$ was
reshaped to $[nx,ny,nz,T]$ to construc the pulsed pressure field $p_\delta$
(\ref{eq:press_field_proportionality}). Five temporal samples are
shown in Figure \ref{fig:simulation_comparison} a. A reduced extent was then located
around the focal point, and the monochromatic pressure  
was computed from $h_{sir}$ following (\ref{eq:monochromatic_pressure_field}) 
and is shown in Figure \ref{fig:simulation_comparison} b-c.

\subsection{RF and transducers PSF simulation}

\subsubsection{Spatio-Temporal Accuracy of RF: The Circular Single-Element PSF case}

A convenient metric for evaluating the spatio-temporal accuracy of the receive
simulation is the computation of the Point Spread Function (PSF) of a transducer. The
PSF describes the response of a focused acoustic beam to an idealized point scatterer.
To assess the accuracy of the proposed receive model, the Field II PSF
example \cite{jensen_jorgen_arendt_field_1996} was reproduced using the receive simulation methods
discussed here and implemented in eSDIva. The reference Field II example computes the
PSF of a focused single-element transducer
by laterally scanning a point scatterer from ($-10\,\mathrm{mm}$) to ($10\,\mathrm{mm}$) in
steps of ($0.2\,\mathrm{mm}$), while maintaining a fixed axial distance of ($30\,\mathrm{mm}$)
from the transducer aperture. The transducer is modeled as a concave circular element
with a radius of ($8\,\mathrm{mm}$), a geometric focus of ($80\,\mathrm{mm}$), and a center
frequency of ($3\,\mathrm{MHz}$). \\

\begin{figure*}[!t]
\centerline{\includegraphics[width=\textwidth]{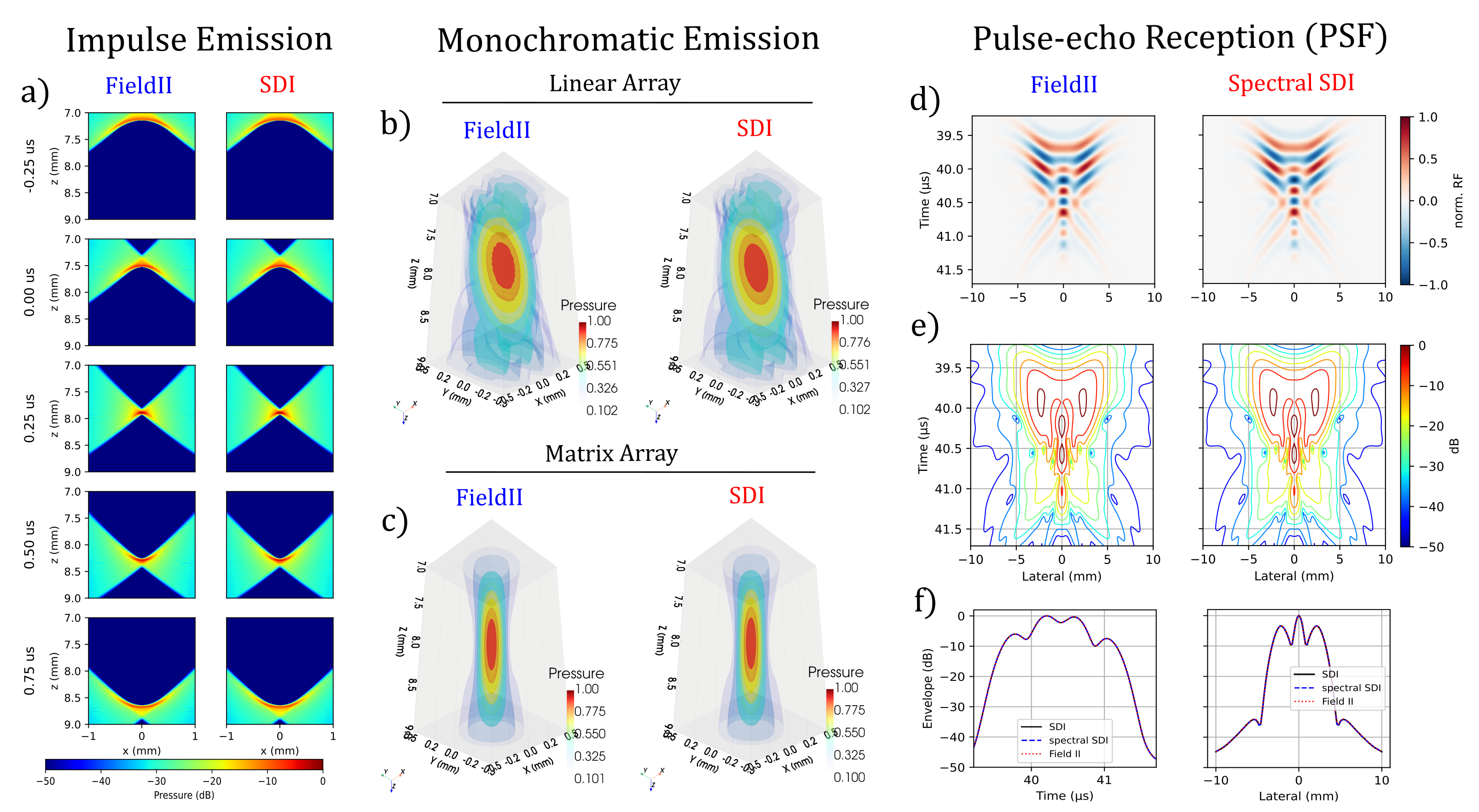}}
\caption{\textbf{Comparison of acoustic field and pulse-echo simulations obtained with
    the proposed SDI method and Field II.}
\textbf{a)} Comparison of $xz$-plane impulse-response simulations for a linear array
focused at $(0,0,8)$~mm without apodization.
\textbf{b)} Comparison of 3D monochromatic pressure fields for a linear array with
elevation focusing and electronic focusing at $(0,0,8)$~mm, without apodization.
\textbf{c)} Same as (b), but for a flat matrix array with Hamming apodization ($F/D =
1$).
\textbf{d)} Comparison of the pulse-echo RF response (point spread function, PSF) of a
focused concave single-element transducer for an ideal point scatterer, simulated with
Field II and SpectralSDI. The horizontal axis denotes the scatterer position, and the
vertical axis the corresponding received RF signal.
\textbf{e)} Envelope of the PSF shown in (d), log-compressed to a 50~dB dynamic range
with 6~dB contour intervals.
\textbf{f)} Axial and lateral profiles of the PSF envelope shown in (e), extracted
through the peak-amplitude location.
}
\label{fig:simulation_comparison}
\end{figure*}

For each scatterer position, the received RF signal was simulated. Since the scan
consists of only 100 scatterer locations, the computational cost is negligible for all
simulation platforms considered. The envelope of each RF signal was then computed and
log-compressed to a dynamic range of ($50\mathrm{dB}$). Contour maps with
($6\mathrm{dB}$) spacing are shown in Figure~\ref{fig:simulation_comparison}  to assess
the spatio-temporal agreement between Field II, the conventional pulse-echo computation
with the SDI method for SIRs (\ref{eq:RF_conventional_signal}), and the proposed
spectral SDI approach (\ref{eq:RF_spectral_SDI}).\\

\subsubsection{Linear and Matrix Array RF Simulation times}

To evaluate the accuracy of the proposed methods beyond the PSF level and up to fully
developed speckle patterns, RF simulations were performed using the same linear and
matrix array transducers employed in the SIR computation benchmarks. Simulations were
conducted for $N_{\mathrm{scat}} \in \{10^2,10^3,10^4,10^5\}$ point scatterers randomly
distributed within an $XZ$ plane with spatial extents $x \in [-2,2]~\mathrm{mm}$ and $z
\in [5,11]~\mathrm{mm}$. All scatterers were assigned a unit reflectivity coefficient,
$\gamma_p = 1$.  \\

For each configuration, the RF signals, scatterer coordinates, and execution time of the
\texttt{calc\_scat\_multi} function in Field II were recorded. The same scatterer
coordinates were then imported into eSDIva, and the execution time was measured from
the start of the RF simulation call until the output RF data were returned. This timing
therefore includes grid organization, computation of the SIRs using either the
time-domain analytical expressions of Eq.~$(\ref{eq:double_int_sir})$ or the
frequency-domain formulation of Eq.~$(\ref{eq:spectral_deltas})$, as well as the forward
and inverse Fourier transform operations required for convolution in the conventional
formulations of Eqs.~$(\ref{eq:RF_conventional_signal})$ and
$(\ref{eq:RF_spectral_SDI})$. \\

For all tested configurations, the channel-wise correlation coefficient between the RF
signals generated by Field II and eSDIva was computed as

\begin{equation}
    \rho = \frac{\mathrm{cov}(RF_{\mathrm{Field II}}, RF_{\mathrm{eSDIva}})}
    {\sigma_{RF_{\mathrm{Field II}}} \sigma_{RF_{\mathrm{eSDIva}}}},
\end{equation}

where $\mathrm{cov}$ denotes the covariance and $\sigma$ the standard deviation of each
signal. Since amplitude scaling might differ between the two implementations, the
correlation coefficient provides a robust measure of similarity between the two RF
signals.

\begin{table}[h]
\newcommand{\cdarkgrey}{\cellcolor[rgb]{.851,.851,.851}}
\newcommand{\clightgrey}{\cellcolor[rgb]{.949,.949,.949}}
\newcommand{\cred}{\cellcolor[rgb]{ .961,  .71,  .71}}
\newcommand{\corange}{\cellcolor[rgb]{ .961,  .875,  .663}}
\newcommand{\cgreen}{\cellcolor[rgb]{ .855,  .949,  .816}}

\centering
\caption[Pulse-echo SIR computation time relative to Spectral SDI]{Comparison of
pulse-echo (reception) SIR computation time for linear and matrix transducers varying
the number of scatterers. 
}

\label{tab:tableIII}

\footnotesize
\setlength{\tabcolsep}{3pt}
\renewcommand{\arraystretch}{1.35}
\begin{tabular}{w{c}{3.2em} w{c}{9em} w{c}{3.0em} w{c}{3.0em} w{c}{3.0em} c}

\cmidrule{3-5}
\multicolumn{2}{c}{} &
\multicolumn{3}{c}{\cdarkgrey\textbf{\shortstack{Conventional method time /\\ Spectral SDI time}}} &
\\

\cmidrule{1-5}
\multicolumn{1}{w{c}{3.2em}}{\cdarkgrey\textbf{$N_{\mathrm{scat}}$}} &
\multicolumn{1}{w{c}{9em}|}{\cdarkgrey\textbf{Spectral SDI t (s)}} &
\cdarkgrey\textbf{FST} &
\cdarkgrey\textbf{SDI} &
\cdarkgrey\textbf{Field II} &
\\

\cmidrule{1-6}
 100 & \cgreen 0.037 $\pm$ 0.001 &
\corange 2.27 & \corange 2.31 & \cred 2.35 &
\multirow{5}{*}{\rotatebox{90}{\textbf{Linear Array}}} \\

 1000 & \cgreen 0.258 $\pm$ 0.007 &
\corange 2.57 & \corange 2.54 & \cred 3.26 & \\

 10000 & \cgreen 2.317 $\pm$ 0.005 &
\corange 2.70 & \corange 2.62 & \cred 3.68 & \\

 100000 & \cgreen 21.321 $\pm$ 0.038 &
\corange 2.78 & \corange 2.90 & \cred 4.09 & \\

\cmidrule{1-5}
\multicolumn{5}{c}{RF correlation: $0.9916 \pm 0.0024$} & \\

\cmidrule{1-6}
 100 & \cgreen 1.834 $\pm$ 2.129 &
\corange 4.61 & \corange 4.16 & \cred 4.51 &
\multirow{5}{*}{\rotatebox{90}{\textbf{Matrix Array}}} \\

 1000 & \cgreen 5.501 $\pm$ 0.034 &
\corange 4.72 & \corange 4.44 & \cred 15.44 & \\

 10000 & \cgreen 50.386 $\pm$ 0.324 &
\corange 4.03 & \corange 3.81 & \cred 15.98 & \\

 100000 & \cgreen 421.692 $\pm$ 0.000 &
\corange 4.61 & \corange 4.57 & \cred 19.55 & \\

\cmidrule{1-5}
\multicolumn{5}{c}{RF correlation: $0.9999 \pm 0.0003$} & \\

\cmidrule{1-6}
\end{tabular}

\smallskip
\footnotesize
$N_{\mathrm{scat}}$: number of point scatterers. Times are the mean RF-signal computation
time; ratios give each conventional pulse-echo method (FST, SDI, Field II)
normalized by Spectral SDI. For instance, at $N_{\mathrm{scat}}=$ $100$k for the matrix array
Field II takes $19.55\times421.69\approx8244\,\mathrm{s}$ $\approx2.3\,\mathrm{h}$ against
Spectral SDI's $421.69\,\mathrm{s}\approx7\,\mathrm{min}$; all ratios exceed 1, i.e.\ every
conventional method is slower than Spectral SDI but faster than Field II. 
RF correlation is the Pearson correl. between the methods and Field II RF lines.

\end{table}

\section{Results}\label{sec:results}

\subsection{Linear and Matrix Array SIR and Emitted Pressure}

\subsubsection{SIR Computation Time and Accuracy}

Table~\ref{tab:tableII} reports the values of $\overline{\Delta k}_{m,p}$ and $8+2T/M$,
which characterize the expected performance difference between the FST and SDI methods
for the tested array transducers. The linear array exhibits lower values of
$\overline{\Delta k}_{m,p}$ and higher values of $8+2T/M$ than the matrix array,
resulting in a smaller performance advantage of SDI over FST, with speedups ranging from
approximately $1.2\times$ to $1.5\times$. \\

As predicted by Eq.~(\ref{eq:final_heuristic_condition}), SDI provides the largest
performance gains for large apertures, explaining the greater acceleration observed for
the matrix array. In this configuration, SDI achieves speedups ranging from
approximately $2\times$ to nearly $4\times$ relative to FST at high sampling
frequencies. The advantage decreases for smaller apertures and with increasing
subdivision counts, as smaller patches reduce the relative benefit of SDI. \\

Across all configurations shown in Table~\ref{tab:tableII}, SDI achieved speedups
ranging from approximately $20\times$ to $180\times$ relative to Field II while
maintaining high numerical accuracy, with MSE values below $10^{-6}$. \\ 

\subsubsection{Emitted Pressure Visualization}

Visual inspection of the transient pulsed-pressure fields shows agreement with Field II
down to approximately $-50\mathrm{dB}$ throughout the simulated time sequence, as
illustrated in Fig.~\ref{fig:simulation_comparison}a. This agreement also extends to the
monochromatic 3D pressure fields shown in Fig.~\ref{fig:simulation_comparison}b-c. for both
linear and matrix array geometries, where no visible differences are observed between
the reconstructed fields. These results further support the accuracy of the proposed SIR
computation in both the time and frequency domains.

\subsection{RF and Transducer PSF Simulation}

\subsubsection{Spatio-Temporal Accuracy of RF: Circular Single-Element PSF}

Figure~\ref{fig:simulation_comparison}d. shows the RF signals used to compute the
PSF of the focused circular single-element transducer using Field II and Spectral SDI.
The simulated RF waveforms are indistinguishable, with a Pearson correlation coefficient
of 1 for all scatterer positions.\\

The corresponding envelope-detected PSFs, displayed over a dynamic range of
$50\mathrm{dB}$ in Fig.~\ref{fig:simulation_comparison}e., also show close
agreement. This is further illustrated in Fig.~\ref{fig:simulation_comparison}f.,
where the lateral and on-axis axial envolepe profiles obtained with Spectral SDI,
conventional SDI, and Field II overlap. These results indicate that the proposed receive
simulations reproduce the Field II reference solution with high spatio-temporal
accuracy.\\

\subsubsection{Linear and Matrix Array RF Simulation Times}

Two RF simulation strategies were evaluated: a conventional approach, in which the
temporal SIR is first computed using FST or SDI and subsequently
convolved with the excitation in the Fourier domain, and Spectral SDI, which directly
evaluates the analytical frequency-domain formulation.\\ 

Table~\ref{tab:tableIII} compares the execution times of the different RF simulation
methods. The execution times of the conventional FST and SDI implementations are similar
for both array geometries. All conventional eSDIva methods 
outperform Field II, with speedups reaching approximately $5\times$ for the matrix
array.\\ 

The largest performance gains are obtained with Spectral SDI. For the matrix array,
Spectral SDI achieves speedups exceeding $4\times$ relative to the conventional eSDIva
methods and more than $20\times$ relative to Field II. Smaller gains are observed for
the linear array. \\

For all tested configurations, the RF signals generated by the eSDIva methods (Spectral
SDI and conventional FST/SDI) exhibited Pearson correlation coefficients
close to 1 when compared with Field II. The matrix-array RF waveforms were effectively
identical, while small discrepancies were observed for the linear-array configuration
incorporating an elevation lens.

\section{Discussion \& Perspectives} \label{sec:discussion}

\subsection{SDI for SIR computation}

The simulations confirm the theoretical performance model derived in
Eq.~(\ref{eq:heuristic_condition}). The measured distributions of
$\Delta k_{m,p}$ for linear and matrix arrays follow the trends predicted by the
analytical formulation. SDI accurately reconstructs both patch-level trapezoidal SIRs
and the resulting transducer responses, with pressure fields matching those obtained
with Field II. As predicted, the largest gains are obtained for large apertures and
high sampling frequencies, where FST requires a larger number of temporal samples.

\begin{figure}[h!]
\centerline{\includegraphics[width=\columnwidth]{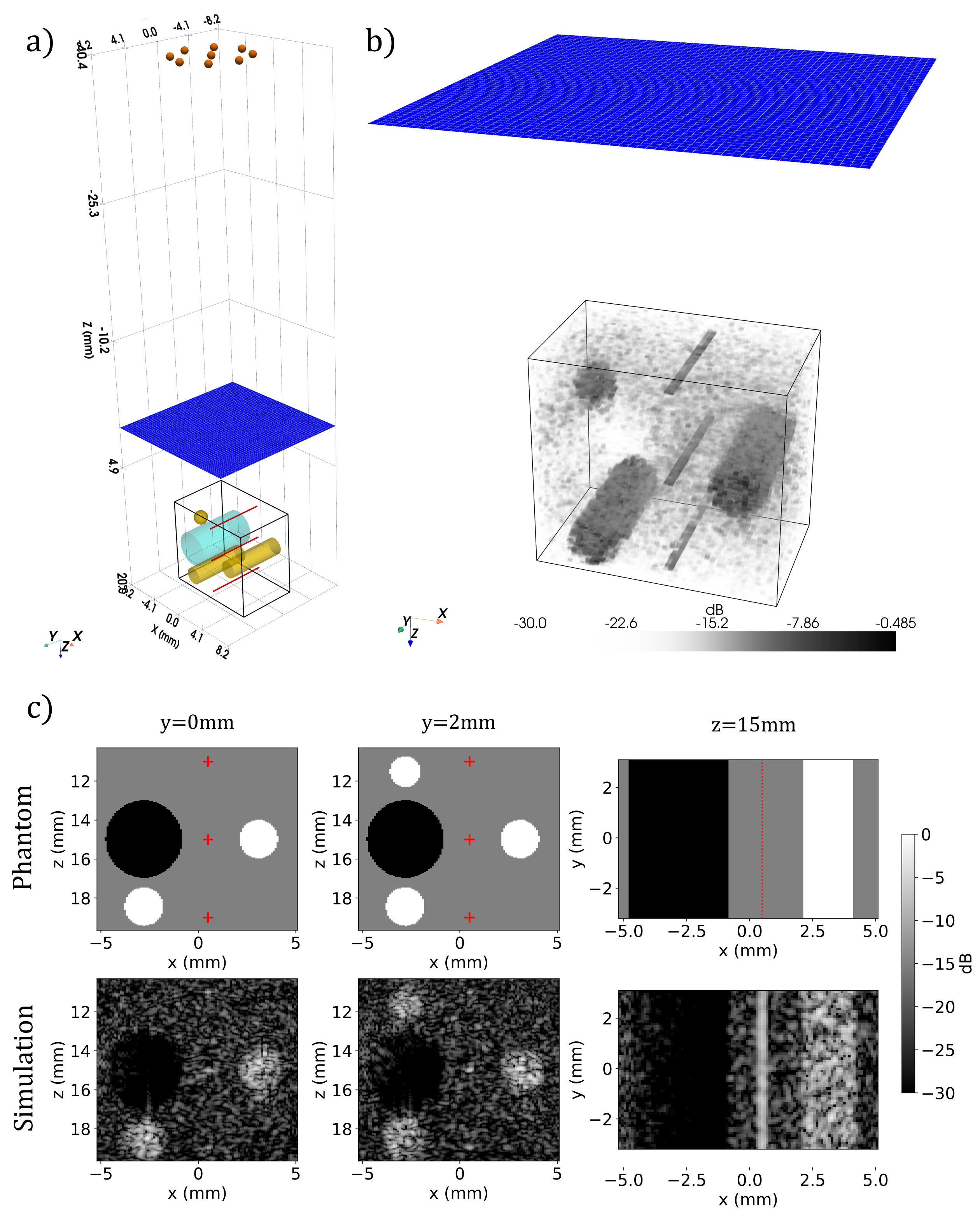}}
\caption[Large-scale phantom simulation using Spectral SDI]{\textbf{Large-scale phantom simulation using Spectral SDI and a $55\times55$
matrix array operating at 10~MHz.}
To generate fully developed speckle, approximately 5--10 scatterers are required per
resolution cell \cite{jensen_fast_2000}, whose volume is approximated by
$V_{\mathrm{cell}}\approx\lambda^3(z/D)^2$. For the simulated phantom
($x\in[-5.5,5.5]$mm, $y\in[-3.5,3.5]$mm, $z\in[10,20]$mm), using five scatterers
per resolution cell resulted in approximately $1.3\times10^6$ point scatterers. A
sequence of nine diverging-wave transmissions was simulated, beamformed using
delay-and-sum (DAS), and coherently compounded. Each transmit event required
approximately 41~min with Spectral SDI, yielding a total simulation time of
$\sim$6h, compared with an estimated $\sim$250-350h (various days time-scale) using
Field II on the same hardware.\footnotemark
\textbf{(a)} Simulation setup showing the nine virtual sources (orange) behind the
matrix array and the phantom containing wire targets (red), anechoic cysts (cyan),
and hyperechoic inclusions (yellow).
\textbf{(b)} Beamformed and compounded volumetric B-mode image.
\textbf{(c)} Comparison between the phantom reflectivity map and the corresponding
simulated B-mode slice.}
\label{fig:phanton_simulation}
\end{figure}

\footnotetext{Estimated from simulations with
$N_{\mathrm{scat}}=\{200,500,1000,5000\}$, which showed an approximately constant
processing time of $\sim9\times10^{-2}$~s per scatterer for this transducer using
Field II.}

\subsection{SDI formulations for reception}

The paired SDI formulation of Eq.~(\ref{eq:RF_SDI_paired}) was not adopted in the final
implementation because its computational cost scales as
$M_{\mathrm{tx}}M_{\mathrm{rx}}$. Nevertheless, it illustrates the flexibility of the
SDI framework for deriving alternative analytical formulations, as exemplified by
Spectral SDI.\\

RF benchmarks show that faster SIR computation alone does not directly translate into
proportional RF speedups, since the conventional pulse-echo pipeline is dominated by
Fourier-domain convolutions. Consequently, the FST and SDI implementations
exhibit similar execution times. In contrast, Spectral SDI derives the SIR directly in
the frequency domain, eliminating part of the convolution pipeline and providing the
largest acceleration. Unlike frequency-domain solvers such as SIMUS, which rely on
paraxial time-harmonic approximations \cite{garcia_simus_2022,cigier_simus_2022},
Spectral SDI operates in the Fourier domain while preserving the exact far-field SIR, so
the computed field is not subject to paraxial error. \\

The computational efficiency of Spectral SDI enables simulations that are otherwise
impractical on a single workstation. For example,
Fig. \ref{fig:phanton_simulation} presents a large-scale simulation of a
$55\times55$ matrix array operating at 10MHz with approximately
$1.3\times10^6$ scatterers. The complete nine-transmit acquisition was simulated in
approximately $6.2$h, compared with an estimated $\geq$250h ($\sim 10$ days) using Field
II on the same hardware.\\

Across all configurations, Pearson correlation coefficients remain close to unity,
confirming that these computational gains are achieved without compromising RF
accuracy. The small discrepancies observed for the linear array are attributed to
differences in elevation-focusing implementation: eSDIva employs geometric delays,
whereas the corresponding Field II implementation is not publicly available. The
agreement obtained for the concave single-element transducer further supports the
validity of this approximation.

\subsection{Computational structure and implementation}

A key advantage of SDI is its fixed computational structure. Each patch contribution is
represented by eight temporal samples independently of aperture geometry or field-point
location, whereas FST requires sampling the full trapezoidal support, which varies with
each patch--field-point pair. This fixed pattern enables efficient vectorization and
provides a natural path toward GPU implementations.\\

Similarly, Spectral SDI provides an analytical frequency-domain SIR expression,
allowing direct integration of frequency-dependent effects such as attenuation while
maintaining the same reusable computational structure. Additional performance gains in
eSDIva result from implementation-level optimizations: the Python FST implementation
combined with Numba-based parallelization already achieves speedups of approximately
$20\times$ in emission and up to $4\times$ in reception compared with Field II.

\section{Conclusion}

We introduced Sparse Delta Integration (SDI), a method for computing far-field spatial
impulse responses (SIRs) of rectangular apertures. SDI reformulates the trapezoidal SIR
as a sum of delta distributions, resulting in a fixed eight-sample representation
independent of field point, aperture geometry, or patch position. In contrast to
Fully Sampled Trapezoid (FST)-based approaches, whose computational cost scales with
the temporal support of the SIR, SDI maintains a constant computational structure.\\

A heuristic condition was derived to predict the regimes in which SDI outperforms FST.
SDI is advantageous when the maximum aperture dimension
$w=\max(w_x,w_y)$ exceeds the critical value $w_{crit}=8c_0/f_s$, corresponding to cases in
which the FST representation requires more than eight samples. Numerical results
confirm this prediction, with SDI providing algorithmic speedups of up to
$\sim4\times$ over FST for SIR accumulation.\\

For pulse-echo simulations, we further derived the Spectral SDI formulation, which
provides an exact frequency-domain representation of the SIR. By eliminating two
forward Fourier transforms from the conventional convolution pipeline, Spectral SDI
substantially accelerates RF simulation while preserving the numerical equivalence of
the underlying SIR formulation.\\

When implemented in the open-source eSDIva framework, these algorithmic improvements
translate into substantial end-to-end performance gains relative to Field II. For
SIR simulations, eSDIva achieves speedups of up to
$\sim 180\times$, while pulse-echo RF simulations using Spectral SDI achieve speedups
of up to $\sim 20\times$. Validation against Field II demonstrates excellent numerical
agreement across all configurations, with mean squared errors below $10^{-5}$ and
Pearson correlation coefficients close to unity.\\

Finally, eSDIva provides an open-source, memory-efficient, and Python-native
simulation framework supporting arbitrary transducer geometries, electronic focusing,
apodization, pressure-field computation, and pulse-echo RF simulation. Its modular
architecture facilitates reproducible research, large-scale ultrasound simulations,
and future integration with optimization, inverse problems, and differentiable
computational imaging methods.

\appendix
\titleformat{\section}{\normalfont\large\bfseries}{\appendixname~\thesection:}{0.5em}{}

\section{Theoretical performance between FST and SDI}
\label{appendix:heuristic_condition}

\subsection{SIRs heuristic inequatility for SDI gain over FST}
The number of trapezoids to be accumulated over the temporal grid is equal to the number of combinations of 
field points and patches on the transducer. In the FST approach, each patch $m$ is evaluated by explicitly 
filling the trapezoid defined by the piecewise function in (\ref{eq:trapezoid_sir}) over
all its nonzero $k$-temporal samples. If, on average, the trapezoids span $\overline{\Delta k}$
sample times over a temporal grid of $T$ samples, then for $P$ field points the computational time (CT) scales as

\begin{equation}
\mathrm{CT}_{\mathrm{FST}}
\sim
\mathcal{O}\!\left(P \, M \, \overline{\Delta k}\right).
\label{eq:ct_naive}
\end{equation}

For clarity, consider the case of a single field point ($P = 1$):

\begin{equation}
\mathrm{CT}_{\mathrm{FST},\,P=1}
\sim
\mathcal{O}\!\left(M \, \overline{\Delta k}\right).
\label{eq:ct_naive_p1}
\end{equation}

In contrast, the Sparse Delta Integration (SDI) approach introduced in
(\ref{eq:double_int_sir}) avoids explicitly filling entire trapezoids. Instead, each
patch contributes only eight Dirac delta functions (\ref{eq:delta_decomposition})
as illustrated in Fig. \ref{fig:rectang_sir}, resulting in a cost
$\mathrm{CT}\left(\sum_{m=1}^{M} \Delta\delta(t-\tau_m)\right) 
\sim \mathcal{O}(8M)$, at the expense of performing two time integrations, each
requiring approximately $\mathcal{O}(T)$ operations. Hence,

\begin{equation}
\mathrm{CT}_{\mathrm{SDI},\,P=1}
\sim
\mathcal{O}(8M + 2T).
\label{eq:ct_SDI_p1}
\end{equation}

The SDI method is advantageous when its computational cost is smaller than that of the FST sample-wise approach:

\begin{equation}
M \, \overline{\Delta k} \gg 8M + 2T,
\label{eq:ct_comparison}
\end{equation}

Which leads to

\begin{equation}
\overline{\Delta k} \gg 8 + \frac{2T}{M}.
\label{eq:heuristic_condition_appendix}
\end{equation}

Note that this condition also holds for the general case $P \neq 1$.\\

\subsection{SIRs heuristic inequality in terms of physical parameters}

The discrete trapezoid width $\Delta k_{m,p}$ for patch $m$ at field point $p$ is given
by

\begin{equation}
\Delta k_{m,p} = f_s \cdot (\Delta t_1 + \Delta t_2)=f_s\cdot \left(\frac{w_x |u_x|}{c_0}+\frac{w_y |u_y|}{c_0}\right)
\label{eq:delta_k_mp}
\end{equation}

Where $f_s$ is the sampling frequency. Assuming $w_x \approx w_y \equiv w$, this simplifies to

\begin{equation}
\Delta k_{m,p} \approx \frac{f_s \cdot w}{c_0}\cdot \left(|u_x| + |u_y|\right).
\label{eq:delta_k_mp_simplified}
\end{equation}

If we imagine field points uniformly distributed in the half‑space $z\geq 0$, then the 
vectors drawn from the patch center to those points define a uniform distribution of 
directions over the upper hemisphere. Then the average value of $|u_x| + |u_y|$ evaluates to:

\begin{equation}
    \frac{1}{2 \pi} \int_{u_z \geq 0} (|u_x| + |u_y|) d\Omega =  1
\label{eq:hemispherical_avg}
\end{equation}

Where $d\Omega$ denotes the solid-sangle element. Therefore, the average trapezoid width
across all patches and directions can be approximated as

\begin{equation}
\overline{\Delta k} \approx \frac{f_s\cdot w}{c_0}.
\label{eq:mean_delta_k}
\end{equation}

A conservative estimate of the temporal bounds is $t_{\mathrm{start}} \approx \frac{l_{\min}}{c_0} - \frac{w}{c_0},
 t_{\mathrm{end}} \approx \frac{l_{\max}}{c_0} + \frac{w}{c_0}$, so that the total temporal span becomes

\begin{equation}
\Delta t_{\mathrm{span}} \approx \frac{l_{\max} - l_{\min} + 2w}{c_0}
\approx (1+\varepsilon)\cdot\frac{l_{\max}}{c_0}
\label{eq:time_span}
\end{equation}

where $\varepsilon = (2w - l_{\min})/l_{\max}$ and tends to zero when $l_{\max} \gg l_{\min}$. The corresponding 
number of time samples is therefore

\begin{equation}
T \approx (1+\varepsilon)\cdot \frac{l_{\max} f_s}{c_0}.
\label{eq:num_time_samples}
\end{equation}

The total number of patches can be estimated as

\begin{equation}
M = \frac{A}{w_x w_y} \approx \frac{A}{w^2},
\label{eq:num_patches}
\end{equation}

Where $A$ denotes the transducer aperture area. Substituting 
(\ref{eq:heuristic_condition_appendix}), (\ref{eq:mean_delta_k}), (\ref{eq:num_time_samples}), and (\ref{eq:num_patches}) yields

\begin{equation}
\frac{f_s w}{c_0} \gg 8 + 2\cdot(1+\varepsilon)\cdot\frac{l_{\max} w^2}{A}\cdot\frac{f_s}{c_0}.
\label{eq:heuristic_substituted_appendix}
\end{equation}

Which leads to the next inequality

\begin{align}
A \gg 2 \cdot \frac{1+\varepsilon}{1 - w_{crit} / w}\cdot w \cdot l_{\max} 
\label{eq:final_heuristic_condition_appendix}
\end{align}

Subject to a patch-size dimensions that satisfies the next condition:

\begin{equation}
w > w_{crit} = \frac{8 c_0}{f_s},
\label{eq:critical_patch_size_appendix}
\end{equation}

\section{Discretization of time} \label{appendix:discretization}

The continuous variable $t$ is replaced by its sampled counterpart $\{t_k\}$, a vector
of $T$ samples defined as

\begin{equation}
t_k = t_{\mathrm{start}} + k \Delta t, \qquad k = 0,1,2,\ldots,T-1,
\label{eq:temporal_grid}
\end{equation}

with sampling step $\Delta t = 1 / f_s$, where $f_s$ is the sampling frequency. The values $t_{\mathrm{start}}$ and 
$t_{\mathrm{end}}$ define a temporal interval large enough to contain all patch SIRs. The total number of temporal 
samples required is then

\begin{equation}
T = \left \lceil(t_{\mathrm{end}} - t_{\mathrm{start}})\cdot f_s \right \rceil
\label{eq:nu m_temporal_samples}
\end{equation}

\section{SDI analytic expression for Reception}
\label{appendix:SDI_pulse_echo}

The conventional formulation of Section~\ref{sec:reception} first builds the two one-way
spatial impulse responses, the emission SIR $h^{e}$ and the reception SIR $h^{r}$, and
then evaluates the received pressure (\ref{eq:received_pr}) by convolvieg them. The
Sparse Delta Integration (SDI) method offers an alternative route: instead of
constructing the two SIRs explicitly and convolving them, it assembles the \emph{two-way}
(pulse-echo) SIR directly from a small set of Dirac deltas. This subsection derives that
analytic expression step by step and shows that it can be evaluated in two equivalent
ways, which we call the \emph{paired SDI} form and the \emph{spectral SDI} form. \\

The pulse-echo SIR is the convolution of the emission and reception SIRs. Substituting
(\ref{eq:double_int_sir}) for each one-way SIR and using the addition and communation of
the integration operator in convolution ($I^2$ applied twice is $I^4$), we obtain

\begin{align}
    h_{pe}(t)
    &= h^{e}(\mathbf{r}_{m_e,p},t) \overset{t}{*}
       h^{r}(\mathbf{r}_{p,m_r},t) \\
    &= \big[I^2 D^2 h^{e}\big] \overset{t}{*} \big[I^2 D^2 h^{r}\big] \\
    &= I^4 \big[\, D^2 h^{e}(\mathbf{r}_{m_e,p},t) \overset{t}{*}
                   D^2 h^{r}(\mathbf{r}_{p,m_r},t) \,\big] \\
    &= I^4 \big[\, \Delta\delta^{e}(\mathbf{r}_{m_e,p},t) \overset{t}{*}
                   \Delta\delta^{r}(\mathbf{r}_{p,m_r},t) \,\big]
       \label{eq:conv_deltas_reception} \\
    &= I^4\, \Delta\delta_{pe}(\mathbf{r}_{m_e,p}, \mathbf{r}_{p,m_r}, t).
       \label{eq:h_pe_SDI}
\end{align}

The decisive simplification is that the convolution of two Dirac deltas is again a Dirac
delta, shifted to the sum of their positions: $\delta(t-a) \overset{t}{*} \delta(t-b) =
\delta(t-a-b)$. Applying this to the convolution in (\ref{eq:conv_deltas_reception})
turns the four emission deltas and four reception deltas into a single sparse two-way
delta train,

\begin{multline}
    \Delta\delta_{pe}(\mathbf{r}_{m_e,p}, \mathbf{r}_{p,m_r}, t) = \\
    s_{m_e,p}\, s_{p,m_r} \sum_{i=1}^{4}\sum_{j=1}^{4} \sigma_i \sigma_j\,
    \delta\!\big(t - t^e_i - t^r_j\big).
    \label{eq:delta_pe}
\end{multline}

Each emission corner ($i$) pairs with each reception corner ($j$), so a single
transmit-receive patch pair produces $4 \times 4 = 16$ pulse-echo deltas, located at the
summed corner times $t^e_i + t^r_j$ and weighted by the products of the corner slopes
$\sigma_i \sigma_j$. Equation~(\ref{eq:h_pe_SDI}) then states the central result: the
two-way SIR is the \emph{fourth} time-integral of this sparse 16-delta train. The two
double integrations of the one-way SIRs have merged into one fourth-order integration
$I^4$. \\

Inserting the two-way SIR into the received-signal equation (\ref{eq:received_pr})
gives the SDI in reception. Because $I^n$ denotes $n$-fold time integration, which in 
the Fourier domain is simply the division $\div (j\omega)^n$, two equivalent 
computational forms arise by arraging the convolution either in the time domain or in
the Fourier domain. They
produce identical results but distribute the work differently, and they have opposite
cost behaviour. We call them 
the \emph{paired SDI} form (time-domain delta placement, one term per patch pair) and
the \emph{spectral SDI} form (Fourier-domain, transmit and receive contributions
separately).

\subsection{Paired SDI}

Using (\ref{eq:delta_pe}) and (\ref{eq:h_pe_SDI}), the received-signal contribution
(\ref{eq:RF_signal}) of a single scatterer and a single transmit-receive patch pair
($P = M_e = M_r = 1$) reads

\begin{multline}
    RF^{SDI}(p, m_e, m_r, t) = \\
    \gamma_p \left[\, \nu_{pe}(t) \overset{t}{*}
    I^4 \Delta\delta_{pe}(\mathbf{r}_{m_e,p}, \mathbf{r}_{p,m_r}, t) \right].
    \label{eq:pulseecho_SDI_paired}
\end{multline}

Here we move the fourth integration onto the excitation/impulse-response waveform,
defining the integrated pulse-echo excitation $E(t) \equiv I^4 \nu_{pe}(t)$, which is
computed once per excitation. Convolving $E$ with the 16 deltas of
(\ref{eq:delta_pe}) simply places 16 shifted, scaled copies of $E$,

\begin{multline}
    RF^{SDI}(p, m_e, m_r, t) = \\
    \gamma_p\, s_{m_e,p}\, s_{p,m_r} \sum_{i=1}^{4}\sum_{j=1}^{4} \sigma_i \sigma_j\;
    E\!\big(t - t^e_i - t^r_j\big).
    \label{eq:RF_SDI_paired}
\end{multline}

The total signal at receive element $e_{rx}$ is the sum over all scatterers and all
transmit-receive patch pairs,

\begin{equation}
    RF(e_{rx},t) = \sum_{p=1}^{P} \sum_{m_e = 1}^{M_e} \sum_{m_r = 1}^{E_{r}}
    RF^{SDI}(p, m_e, m_r, t).
    \label{eq:RF_signal_SDI_annexes}
\end{equation}

Equations~(\ref{eq:delta_pe}), (\ref{eq:pulseecho_SDI_paired}) and
(\ref{eq:RF_SDI_paired}) define the paired-delta form. For each scatterer $p$ and each
patch pair $(m_e, m_r)$ it places the $16$ deltas of (\ref{eq:delta_pe}) in continuous
time. After discretization (Section~\ref{sec:discrete_implementation}) each delta is
split between its two neighbouring samples by linear interpolation, so each patch pair
costs $16 \times 2 = 32$ sample updates. The form is called \emph{paired} because the
computation iterates over transmit-receive patch \emph{pairs}: its cost therefore grows
with the \emph{product} $M_e M_r$ of the transmit and receive patch counts.

\subsection{Spectral SDI form}

Using (\ref{eq:conv_deltas_reception}) instead, the same single-scatterer, single-pair
contribution is

\begin{multline}
    RF^{SDI}(p, m_e, m_r, t) = \\
    \gamma_p \left[\, \nu_{pe}(t) \overset{t}{*}
    I^4 \big[\, \Delta\delta^{e}(t) \overset{t}{*} \Delta\delta^{r}(t) \,\big] \right].
    \label{eq:pulseecho_SDI_factored}
\end{multline}

We now evaluate this in the Fourier domain. The convolutions become products and the
integration operator $I^4$ becomes the division $1/(j\omega)^4$,

\begin{equation}
    \mathcal{F}\{RF\}_{(\omega)} = \frac{\gamma_p}{(j\omega)^4}\,
    V_{(\omega)} \times \mathcal{F}\{\Delta\delta^e\}_{(\omega)}
    \times \mathcal{F}\{\Delta\delta^r\}_{(\omega)},
    \label{eq:FT_pulseecho_SDI_factored}
\end{equation}

where $V_{(\omega)} = \mathcal{F}\{\nu_{pe}\}$. The key advantage appears here: the
Fourier transform of a sparse delta train is a closed-form sum of complex exponentials,

\begin{equation}
    \mathcal{F}\{\Delta\delta^{e/r}\}_{(\omega)} = s_{p,m_{e/r}} \sum_{i=1}^{4} \sigma_i
    \exp\!\big(-j\omega t_i^{e/r}\big),
    \label{eq:FT_delta}
\end{equation}

So no time-domain sampling of the one-way SIR is needed at all. Using
(\ref{eq:TOFs1})-(\ref{eq:TOFs}) to compute the corner times $t_i^{e/r}$, equation 
(\ref{eq:FT_delta}) can be expressed as

\begin{multline}
    \mathcal{F}\{\Delta\delta^{e/r}\}_{(\omega)} = s_{p,m_{e/r}} 
    e^{-j\omega t_1^{e/r}} \\
    \left( 1 -  e^{-j\omega \Delta t_1^{e/r}} \right) 
    \left(1- e^{-j\omega \Delta t_2^{e/r}}\right) 
\end{multline}

\begin{equation}
    = -4s_{p,m_{e/r}}\sin\left(\frac{\omega\Delta t_1^{e/r}}{2}\right)
    \sin\left(\frac{\omega\Delta t_2^{e/r}}{2}\right) e^{-j \omega t_1^{e/r}}
    \label{eq:FT_delta_envelope}
\end{equation}

Equation (\ref{eq:FT_delta_envelope}) is numerically more stable than
(\ref{eq:FT_delta}). When $\Delta t_1^{e/r}$ and $\Delta t_2^{e/r}$ tend to zero, the
coefficients $s_{p,m_{e/r}}$ can become large and require higher accuracy to be sampled
correctly. Because the 
Fourier transform is linear, the sums over the transmit patches and over the receive
patches can therefore be carried out \emph{independently} before they are multiplied.
Summing (\ref{eq:RF_signal_SDI_annexes}) over patches and points and grouping the
transmit and receive contributions gives the factored received signal,

\begin{equation}
    RF(e_{rx},t) = \mathcal{F}^{-1}\Biggl\{ \frac{\gamma_p}{(j\omega)^4}\, V_{(\omega)}
        \times \sum_{p=1}^{P}
        \Big[\, \Sigma_{TX} \times \Sigma_{RX} \,\Big] \Biggr\},
    \label{eq:RF_signal_SDI_TF_annexes}
\end{equation}

with the transmit and receive spectral sums using (\ref{eq:FT_delta_envelope}) as follows

\begin{align}
    \Sigma_{TX} &= \sum_{m_e = 1}^{M_e} 
    \mathcal{F}\{\Delta\delta^{e}\}_{(\omega)} \\
    \Sigma_{RX} &= \sum_{m_r = 1}^{E_r} 
    \mathcal{F}\{\Delta\delta^{r}\}_{(\omega)}
\end{align}

The form is called \emph{spectral} because thanks to the convolution theorem
the transmit sum $\Sigma_{TX}$ ($M_e$ patches) and the receive sum
$\Sigma_{RX}$ ($M_r$ patches) are built separately and only then multiplied. The cost
therefore grows with the \emph{sum} $M_e + M_r$ of the patch counts rather than their
product. Structurally, (\ref{eq:RF_signal_SDI_TF_annexes}) mirrors the conventional
formulation (\ref{eq:RF_conventional_signal}), with the one difference that the one-way
responses are now closed-form delta spectra (\ref{eq:FT_delta}) instead of explicitly
sampled SIRs and no numerical integration is needed.\\

The \emph{paired} and \emph{spectral} forms are mathematically identical: both compute
the same $RF(e_{rx},t)$. They differ only in \emph{where} the fourth integration and the
convolution are evaluated. The paired-delta form places $I^4 \nu_{pe}$ as shifted copies
in the time domain, looping over patch pairs and paying a cost proportional to $M_e M_r$.
The spectral SDI form performs the integration and convolution as a spectral product,
exploiting the additive separability of the two-way delay to reduce the cost to
$M_e + M_r$. Which form is cheaper depends on the patch counts, the time-record length
and the number of scatterers.

\bibliographystyle{unsrt}
\bibliography{Bibliography.bib}

\end{document}